\documentclass[5p,times]{elsarticle}

\usepackage{amssymb}

\usepackage[figuresright]{rotating}

\usepackage{url}
\usepackage{comment}
\usepackage[colorinlistoftodos]{todonotes}
\usepackage{blindtext}
\usepackage{multirow}

\usepackage[T1]{fontenc}
\usepackage[utf8]{inputenc}

\begin{document}

\begin{frontmatter}




\title{Evaluating the impact of the L3 cache size of AMD EPYC CPUs on the performance of CFD applications}

\author[PSNC]{Marcin~Lawenda}
\ead{lawenda@man.poznan.pl}

\author[CUT]{{\L}ukasz~Szustak}
\ead{lukasz.szustak@pcz.pl}

\author[SIE]{László~Környei} 
\ead{laszlo.kornyei@math.sze.hu}

\author[HLRS]{Flavio Cesar Cunha Galeazzo} 
\ead{flavio.galeazzo@hlrs.de}

\author[CUT]{Paweł Bratek} 
\ead{pawel.bratek@pcz.pl}

\cortext[cor1]{Corresponding author.}

\address[PSNC]{Poznan Supercomputing and Networking Center, Jana Pawła II 10, 61-139 Poznań, Poland}

\address[CUT]{Czestochowa University of Technology, Dąbrowskiego 69, 42-201 Częstochowa, Poland}

\address[SIE]{Széchenyi István Egyetem-University of Győr, Győr Egyetem tér 1. tanulmányi ép. B-604, Hungary}

\address[HLRS]{High Performance Computing Center Stuttgart (HLRS), University of Stuttgart, Nobelstraße 19, 70569 Stuttgart, Germany}

\begin{abstract}
In this work, the authors focus on assessing the impact of the AMD EPYC processor architecture on the performance of CFD applications. Several generations of architectures were analyzed, such as Rome, Milan, Milan~X, Genoa, Genoa~X and Bergamo, characterized by a different number of cores (64-128), L3 cache size (256 - 1152 MB) and RAM type (8-channel DDR4 or 12-channel DDR5). The research was conducted based on the OpenFOAM application using two memory-bound models: motorBike and Urban Air Pollution. In order to compare the performance of applications on different architectures, the FVOPS (Finite VOlumes solved Per Second) metric was introduced, which allows a direct comparison of the performance on the different architectures. It was noticed that local maximum performance occurs in the grid sizes assigned to the processing process, which is related to individual processor attributes. Additionally, the behavior of the models was analyzed in detail using the software profiling analysis tool AMD \textmu Prof to reveal the applications' interaction with the hardware. It enabled fine-tuned monitoring of the CPU’s behaviours and identified potential inefficiencies in AMD EPYC CPUs. Particular attention was paid to the effective use of L2 and L3 cache memory in the context of their capacity and the bandwidth of memory channels, which are a key factor in memory-bound applications. Processor features were analyzed from a cross-platform perspective, which allowed for the determination of metrics of particular importance in terms of their impact on the performance achieved by CFD applications.

\end{abstract}


\end{frontmatter}



\section{Introduction}
A comprehensive understanding of the technological advancements shaping modern HPC systems is crucial for developing efficient applications that fully utilize the hardware capabilities. 
One of the fundamental elements of the HPC architecture is the central processing unit (CPU), which undergoes significant modifications with succeeding generations. In order to adapt the operation of the application to the architecture (co-design) and take advantage of the opportunities it offers, it is extremely important to understand the data performance for different problem sizes in the context of data flow.

The article presents the implementation of cutting-edge HPC solutions, focusing on AMD technologies relevant to pilot applications, which are also compared to previous processor generations to demonstrate performance advancement. 

It can be noted that processors are eagerly used in the HPC domain, offering a large number of cores, performance and appropriate memory subsystems with large cache capacity. As an example, we can cite the strongest systems in Europe \cite{TOP500} offered by EuroHPC Joint Undertaking \cite{EuroHPC} to the scientific and commercial community.

The AMD EPYC 9004 series processors represent the 4th generation of AMD EPYC server-class processors. The design of this generation features the AMD Zen 4 microarchitecture of compute cores, the AVX-512 instruction set, large cache memory, and high memory bandwidth to meet the needs of HPC applications. AMD EPYC 9004 series processors offer a variety of configurations with varying numbers of cores, TDPs, frequencies, and cache sizes. 

Simulation tools are commonly used to assess the course of phenomena observed in the surrounding environment, e.g. how air, pollutants, smoke or heat spread in spaces with complex geometry. One of the most popular applications used for this purpose is CFD (Computational Fluid Dynamics). CFD is an intensively developing branch of computational sciences that allows the creation of simulations of fluid-flow phenomena based on the laws governing the movement of fluids. By using dedicated data structures combined with numerical analysis, it is possible to solve problems related to fluid flows. The calculations simulate the flow of the fluid (liquid and gas) and its interaction with the surfaces defined by the boundary conditions. One of the leading CFD applications is OpenFOAM \cite{OpenFOAMwebsite}, which is an open-source framework for the solution of Partial Differential Equations (PDE) using the Finite Volume Method (FVM).

The work deals with two use cases based on OpenFOAM framework: motorBike, a well-known reference application, and Urban Air Pollution, implemented by Széchenyi István Egyetem-University, simulating air flow and pollution dispersion in an urban area. This choice of test applications was consciously made to assess the impact of processing performance from the perspective of two different models and, therefore, slightly different data flow processes, which allow for comparative analysis.

The increasing complexity of models and the need to obtain the highest possible accuracy of results result in an increased demand for computing power and its efficient use. In addition to the traditional algorithmic optimization process, the process of increasing software capabilities is also carried out through a co-design procedure, where the software is adapted to better use the available infrastructure features.

This work focuses on benchmarking of new-generation AMD EPYC processors compared to prior generations. In particular, we explore a series of top-of-the-line AMD EPYC CPUs based on Rome, Milan, Milan~X, Genoa, Genoa~X, and Bergamo architectures (AMD EPYC 7002, 7003 and 9004 series processors). To provide a clear research context, the paper includes a detailed explanation of the application architecture of the new processors and the testing platforms. Special attention is given to the impact of large-cache systems on the parallel performance essential for CFD computational kernels.

Since the work aims to present the features of different generations of processors, a general overview of comparative activities is presented from the perspective of a single node for HPC infrastructures. This made it possible to omit the impact of other factors (e.g. inter-node communication) on the application performance and to highlight the features of systems based on AMD processors (e.g. cache size) along with the findings regarding their profiling results.

The paper is structured as follows. Section 2 provides information on related work in assessing the impact and optimization of cache utilization on application performance. Chapter 3 specifies in detail the infrastructure used for research, focusing on memory-related aspects. The next chapter (4) introduces the application aspect, both in terms of the tools used in tests (compilers, measuring cache parameters) and the description of the framework used to run simulation models. 
Chapter 5 presents models constituting the basis for the research, characterized by various workflows (cache use): motorBike and Urban Air Pollution. Sections 6 and 7 (respectively for motorBike and UAP) present information on the benchmarking results of different CPU architectures in the context of different data sizes reflected by selected CPU metrics along with performance evaluation. 
The chapter (8) analyzes the wide range of performance metrics from a cross-platform perspective with reference to the FVOPS metric. The last, 9th chapter summarizes the research conducted and the results obtained.

\section{Related works}
The CFD simulation imposes the need to process large amounts of data resulting from the size and density of the computational mesh. Both factors are important for the modeled phenomenon and the accuracy of the results \cite{MARTINS2014218}.

Connecting the values achieved in the individual grid nodes with the values in neighbouring nodes implies the need for quick access to the input data necessary to determine the values under the assumed boundary conditions \cite{Moureau2011}.

The effective level of cache usage is influenced by two fundamental aspects: its size (the ability to store a certain amount of data) and the data reading prediction algorithm (data availability on request). This translates into the achieved values on read success (cache hit) or read fail (cache miss), which in turn affects the processing efficiency. Achieving high computational performance requires placing data as close to the computational kernels as possible, ideally by performing efficient prefetching targeting specific cache levels based on dedicated strategies \cite{Hadade2020}.

The implementation of HPC nodes with a significant number of cores, on the one hand, allows for running many processes simultaneously and, on the other hand, imposes requirements on the design of collective MPI operations at the intra-node level. One of the elements that should be taken into account is the cache coherence protocol and its impact on the performance of the designed distributed application algorithms. This topic is widely discussed in scientific literature, proving its importance. Some examples are provided below to illustrate.

In a study \cite{10.1145/3605573.3605616} the authors discuss a scenario where the performance of MPI transmission, based on shared memory, is degraded due to the cache coherence protocol and the multi-socket configuration of the tested platforms. To address this issue, an innovative approach is proposed to reduce the delay of broadcast operations to 1.5 times and implement them in a modified version of the MPI transmission that supports cache coherence.

In another work \cite{SZUSTAK2023623} the impact of new generation non-uniform memory access (NUMA) on the performance of social sciences applications is elaborated. Performance analysis of an application simulating the structure of a user interaction graph revealed communication limitations that impacted its overall performance. The solution was to implement an improved method of distributing data to ccNUMA domains, which reduced intra-processor data traffic and enabled more efficient use of available cores.

Another approach to optimizing workload in CFD application (ADflow), based on the cache-aware improvements on modern vector processors is presented in the work \cite{10.1145/3468267.3470615}. It involves dividing computational blocks into smaller, fixed-size blocks that are small enough to completely fit into the available cache size for each core in a given architecture. This enabled the use of modern vector instruction sets like AVX2 and AVX512, achieving as much as 3.27 times acceleration of the basic CFD solver procedures.

\section{Architecture of AMD EPYC CPUs}
\label{sec:cpu}


This work explores a series of top-of-the-line AMD EPYC CPUs with various architectures. In particular, we use seven dual-socket platforms, including four 64-core CPUs with Rome, Milan, Milan~X, and Genoa architectures, two 96-core CPUs with Genoa and Genoa~X processors, as well as one 128-core CPU based on Bergamo processors. The specifications of these systems are presented in Table \ref{tab:platforms}.

\begin{table}[!h]
\caption{Specification of target platforms}
\label{tab:platforms}
\begin{center}
\begin{tabular}{|c|c|c|}
\hline
Platform & CPU & Memory  \\ \hline
GIGABYTE    & 2 $\times$ AMD EPYC & 512GB \\ 
H262-Z63-00 &   7742 (Rome)    & DDR4-3200\\ \hline
GIGABYTE    & 2 $\times$ AMD EPYC & 512GB \\ 
H262-Z63-00 &  7763 (Milan)     & DDR4-3200\\ \hline
Asus RS720A & 2 $\times$ AMD EPYC & 512GB \\ 
E11 RS12    &  7773X (Milan~X)    & DDR4-3200\\ \hline
AMD         & 2 $\times$ AMD EPYC & 768GB \\ 
Titanite 4G &  9554  (Genoa)   & DDR5-4800\\ \hline
AMD         & 2 $\times$ AMD EPYC & 768GB \\ 
Titanite 4G &  9654  (Genoa) & DDR5-4800\\ \hline
AMD         & 2 $\times$ AMD EPYC & 768GB \\ 
Titanite 4G &  9684X  (Genoa~X)   & DDR5-4800\\ \hline
AMD         & 2 $\times$ AMD EPYC & 768GB \\ 
Titanite 4G &  9754 (Bergamo)     & DDR5-4800\\ \hline
\end{tabular}
\end{center}
\end{table}

Table  \ref{tab:cpus} summarizes the parameters of AMD EPYC processors family. The studied 2nd, 3rd, and 4th generation of AMD EPYC  include the different numbers of cores and are clocked by the base frequency clocks listed in Table \ref{tab:cpus}. The CPU designs of all the platforms feature the out-of-order execution model and support the frequency boost technology, where the maximum turbo frequency depends on the type and intensity of workload and the number of utilized cores. The simultaneous multithreading (SMT) is turned off for all the systems.

\begin{table*}[!h]
\caption{Specification of AMD EPYC CPUs}
\label{tab:cpus}
\begin{center}
\begin{tabular}{|c|c|c|c|c|c|c|c|}
\hline
CPU type      & 7742 & 7763 & 7773X & 9554 & 9654 & 9684X & 9754 \\ \hline
Launch Date      & 2019 & 2021 & 2021 & 2022 & 2022 & 2023 & 2023  \\ \hline
Codename      & Rome & Milan & Milan~X & Genoa & Genoa & Genoa~X & Bergamo \\ \hline
Cores         & 64 & 64 & 64 & 64 & 96 & 96 & 128 \\ \hline
Freq. (Turbo) & 2.25 (3.4*)  & 2.45 (3.5*)  & 2.20 (3.5*)   &  3.10 (3.75) & 2.40 (3.55) & 2.55 (3.42)  & 2.25 (3.1) \\ \hline
L2 [MB]       & 0.5 & 0.5 & 0.5 & 1 & 1 & 1 & 1 \\ \hline
CCX    &  4-core  &  8-core  &  8-core  &  8-core  &  8-core  &  8-core  & 8-core   \\ \hline
L3 CCX  [MB]      & 16 & 32  & 96 & 32 & 96 & 32 & 16 \\ \hline
CCXs in CCD  &  2$\times$4-core & 1$\times$8-core & 1$\times$8-core & 1$\times$8-core & 1$\times$8-core & 1$\times$8-core & 2$\times$8-core \\ \hline
L3 CCD [MB]  &  32 & 32 & 96 & 32 &32 &96 &32  \\ \hline
CCDs  &  8 & 8 &8 &8 &12 &12 &8  \\ \hline
Total L3 [MB] & 256 & 256 & 768 & 256 & 384 & 1152 & 256 \\ \hline
L3 per core [MB] & 4 & 4 & 12 & 4 & 4 & 12 & 2 \\ \hline
Memory  & 8-channel & 8-channel & 8-channel & 12-channel & 12-channel & 12-channel & 12-channel \\ 
type  & DDR4-3200 & DDR4-3200 & DDR4-3200 & DDR5-4800 & DDR5-4800 & DDR5-4800 & DDR5-4800 \\ \hline
NUMAs & 4 & 4  & 4  & 4  & 4 & 4 & 4 \\ \hline
\multicolumn{8}{l}{*maximum frequency achievable by single core}
\end{tabular}
\end{center}
\end{table*}

\begin{figure*}[!h]
\centering
\includegraphics[width=0.84\linewidth]{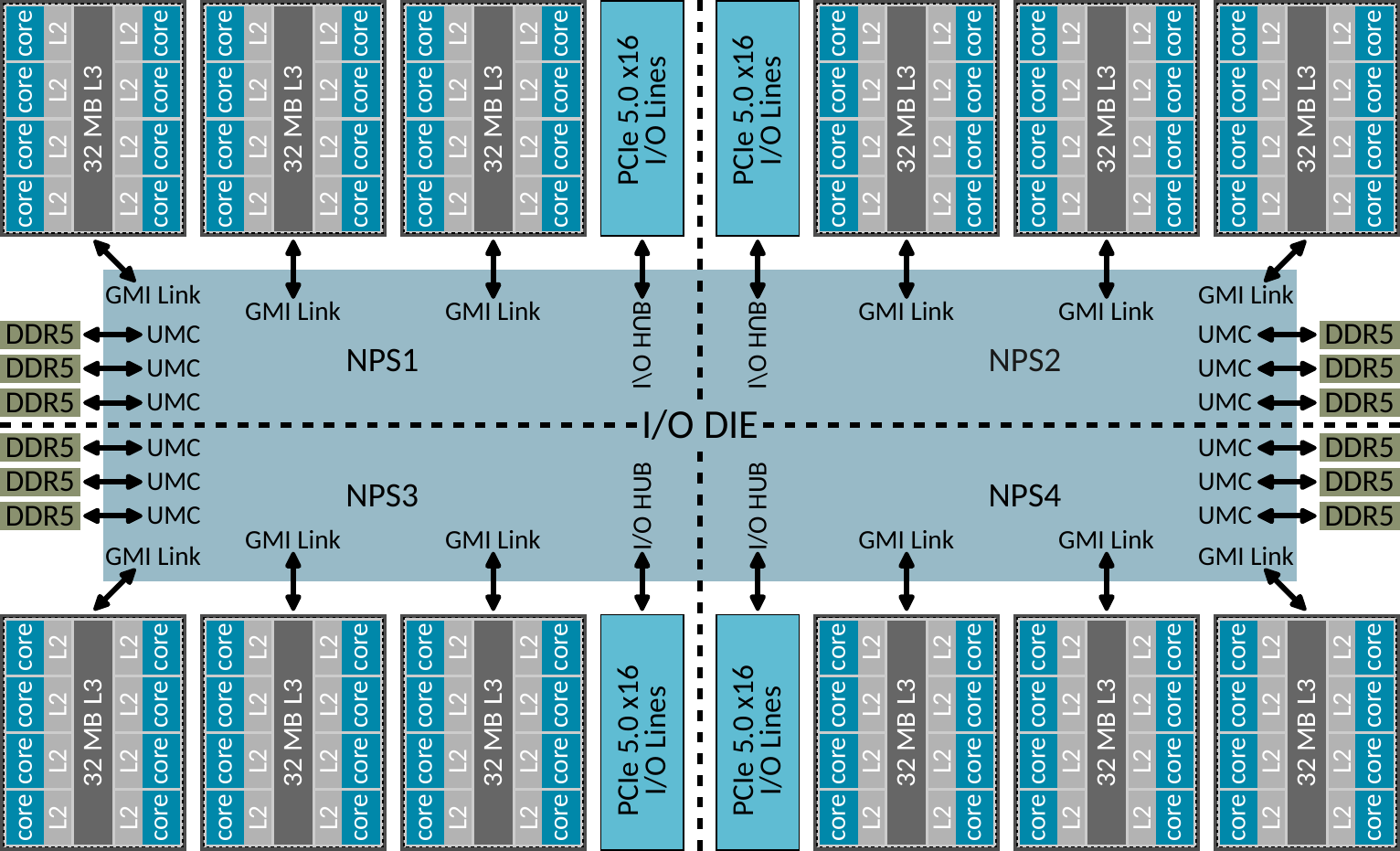}
\caption{AMD EPYC Genoa model with 96 cores, 12 CCDs and a central IOD}
\label{fig:genoa96}
\end{figure*}

The underlined Rome, Milan, and Milan~X chips offer 64-core CPUs and support eight-channel DDR4-3200 main memory per socket. In today’s 4th generation of AMD EPYC processors, we address our work to explore 64-, 96- and 128-core processors that incorporate twelve DDR5-4800 memory channels within every socket \cite{AMD4thGenSpec}. Figure \ref{fig:genoa96} illustrates the block diagram of the 4th generation of AMD EPYC Genoa model with 96 cores. All the platforms represent a group of the ccNUMA shared memory architectures \cite{SZU20_TPDS,SZU21_TPDS} combining whole memory regions using 2x4 NUMA domains for dual-socket platforms (4 domains per socket). A single processor uses four NUMA domains with separate quadrants and, in consequence, interleaves memory regions across the eight and twelve memory channels for the previous (2nd and 3rd) and current 4th generations of the AMD EPYC family, respectively.

The most crucial innovation in AMD EPYC processors is the hybrid multi-die architecture first introduced in 2nd generation of EPYC processors \cite{AMDEPYC2ndGen}. The design of all CPUs consists of a single central I/O hub (or I/O Die) \cite{IOConHub} through which all CPU components communicate. Excluding the Bergamo-based CPUs, the tested AMD EPYC CPUs use a collection of 8-core chiplets, called Core Complex Dies (CCDs), connected to the I/O Die through dedicated high-speed Infinity Fabric links. A single processor with 64-core Rome, Milan, Milan~X, or Genoa chips consists of eight CCDs, while every 96-core Genoa or Genoa~X processor offers configurations with twelve CCDs per socket (see Figure \ref{fig:genoa96}). In contrast, every 128-core Bergamo CPU provides eight CCDs with 16 cores per die.

Every Rome CCD consists of two core complexes (CCXs); each of them embraces four cores and 16 MB of L3 cache. As a result, each CCD provides 32 MB of L3 cache. A single core contains the L2 inclusive cache of 512 KB size and the L1-D cache of 32KB size. The core complex die of the Milan chips holds a single core complex with eight cores and 32 MB of shared L3 cache. Additionally, the configuration of Milan CCDs can be augmented with 3D V-Cache technology to bring the L3 cache capacity to 96 MB in the Milan~X processor. Similarly to Rome, every Milan or Milan~X core offers 512 KB of L2 cache size and the L1-D cache of 32KB size. 

The Genoa CCD used in 64- and 96-core processors consists of one CCX with eight cores, a dedicated 1 MB L2 cache per core, and a 32 MB L3 cache shared between the eight cores. Similarly to Milan~X, applying the 3D V-Cache technology in 4th generation of AMD EPYC CPU enables extending the shared 32 MB L3 cache with 64 MB additional layered above, bringing the per-die total L3 cache to 96 MB in Genoa~X. In contrast to Genoa and Genoa~X, the CCD in Bergamo combines two core complexes with eight cores each. Every core complex includes 16 MB shared L3 cache (32 MB per CCD). 

\begin{figure}[!htb]
\centering
\includegraphics[width=\columnwidth]{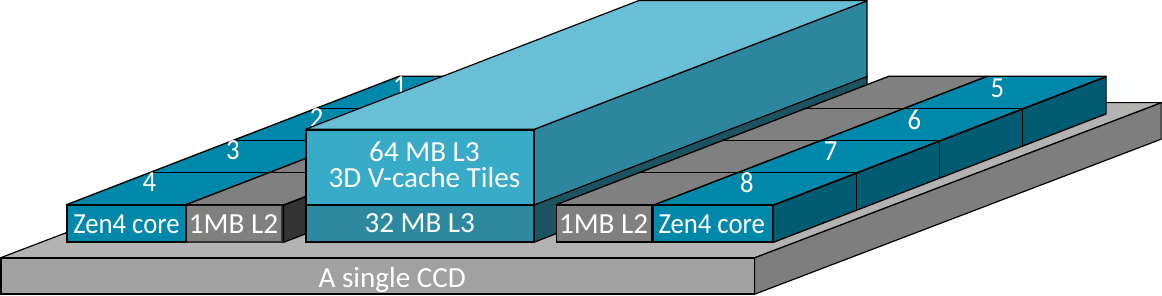}
\caption{Layout of CCD with AMD 3D V-cache enabled}
\label{fig:3Dcache}
\end{figure}

The AMD EPYC processors family typically offers 32 MB of L3 cache per CCD. It results in a total L3 cache capacity of 256 MB for 8 CCDs AMD chips on 64-core Rome, Milan, and Genoa as well as 128-core Bergamo. The 96-core AMD chip contains 12 CCDs and features 384 MB of aggregated L3 cache size. In processors with AMD 3D V-Cache technology, the per-die L3 cache is augmented three times, bringing it to 96 MB per CCD (see Figure \ref{fig:3Dcache}). This innovation delivers a large aggregated L3 cache size that reaches 768 MB for 8 CCDs on a 64-core Milan~X chip and 1152 MB for 12 CCDs on a 96-core Genoa~X chip.

\section{Benchmark applications overview}

\subsection{The motorBike use case} 

The motorBike use case is one of OpenFOAM’s tutorials, that has been with the distribution for a long time. It simulates a motorbike in a wind tunnel and calculates turbulent flow around the vehicle. It showcases the use of snappyHexMesh, OpenFOAM’s own unstructured mesh generation tool, and the simpleFoam solver, which solves the steady-state Reynolds-averaged Navier-Stokes (RANS) equations for an incompressible fluid.

The motorBike use case was chosen for benchmarking and analysis in this report due to its frequent investigation in assessing OpenFOAM performance and its similarities to the UAP-FOAM solver, working with unstructured meshes and solving the RANS equations for incompressible fluid.

The simulation models the external air flow around a motorbike with a pilot in a volume of $10 \times 8 \times 8$ meters. Boundary conditions for bike and ground are modelled as no-slip walls, inlet air speed head-on is set at 20 $m/s$, while side and top boundaries are modelled as slip walls. Meshes of several cell counts for the same geometry are generated using OpenFOAM's snappyHexMesh tool, resulting in mesh sizes from 36k to 14M (see Table \ref{tab:motorMeshes}). All meshes were pregenerated to ensure identical results across all platforms.

\begin{table}[!h]
\caption{Meshes for motorBike test}
\label{tab:motorMeshes}
\begin{center}
\begin{tabular}{|c|c||c|c|c|}
\hline
\multirow{2}{*}{Name} & Cell & \multicolumn{3}{c|}{ Avg. cell count per} \\ \cline{3-5}
& count& 128 cores& 192 cores & 256 cores \\ \hline
xsmall & 167555	 & 1309	& 873	& 655 \\ \hline
small & 355474	 & 2777	& 1851	& 1389 \\ \hline
smid & 603547	 & 4715	& 3143	& 2358 \\ \hline
mid & 1897187	 & 14822	& 9881	& 7411 \\ \hline
high & 4166441	 & 32550	& 21700	& 16275 \\ \hline
mhigh & 6657491	 & 52012	& 34674	& 26006 \\ \hline
uhigh & 11900363 &	92972	& 61981	& 46486 \\ \hline
xhigh & 39400357 & 307815	& 205210	& 153908 \\ \hline
\end{tabular}
\end{center}
\end{table}


The parameters of the simulations are mostly as of in the tutorial use case of OpenFOAM com version v2112. The consistent Semi-Implicit Method for Pressure Linked Equations (SIMPLEC) is used to solve the Reynolds-averaged Navier-Stokes equations for incompressible fluids with k-$\omega$-SST turbulence model. A few adjustments were made to facilitate benchmarking and improve stability. The number of iterations is lowered to 200. The scotch library is used for decomposition \cite{Scotchlib}. The number of decomposed subdomains was equal to the number of available CPU cores on the underlying architecture. Decomposed data was stored in collated format. The number of non-orthogonal correctors was set to one. The underrelaxation factor for velocity was set to 0.7, while for $k$ and $\omega$ to 0.3, and for pressure to $0.3$.

\subsection{The Urban Air Project use case} 
The Urban Air Project (UAP) use case is one of the pilots of the HiDALGO2 project \cite{HiDALGO2}. The CFD module simulates air flow and pollution dispersion in an urban area. In the current paper, the OpenFOAM-based implementation of the CFD module is benchmarked. The UAP workflow gathers data, measurements and simulation results with regard to urban geometry, climate conditions, pollution emission and background, which serve as geometry, boundary conditions and source terms for the CFD model. Two solvers of OpenFOAM, simpleFoam and pimpleFoam \cite{OpenFOAMStanSolv} are utilized to model steady-state and time-dependent behaviour of pollution spread within the city, respectively. In the frame of the current paper only the steady state part is benchmarked using simpleFoam, so results are comparable with the motorBike use case.

The geometry of the use case models external air flow around a geometry resembling the urban area of the Hungarian city of Győr. Boundary conditions use atmospheric wall functions with a roughness factor of $z_0=0.4$ for ground and building, slip for top, and a custom inlet-outlet condition for the side walls that set time and height based values obtained from ECMWF weather service interface, polytope \cite{Polytope}. In the model, pollution is calculated with the COPERT model \cite{Copert} from traffic data obtained by simulation.

\begin{table}[!h]
\caption{Meshes for UAP test}
\label{tab:UAPMeshes}
\begin{center}
\begin{tabular}{|c|c||c|c|c|}
\hline
\multirow{2}{*}{Name} & Cell & \multicolumn{3}{c|}{ Avg. cell count per} \\ \cline{3-5}
& count& 128 cores& 192 cores & 256 cores \\ \hline
uxlow&	36248	&	283	&	189	&	142	\\ \hline
ulow &	139937	&	1093	&	729	&	547	\\ \hline
mlow &	228263	&	1783	&	1189	&	892	\\ \hline
low &	728162	&	5689	&	3793	&	2844	\\ \hline
mid &	3227275	&	25213	&	16809	&	12607	\\ \hline
high &	14332247	&	111971	&	74647	&	55985	\\ \hline
\end{tabular}
\end{center}
\end{table}


Meshes used for these benchmarks are octree based and are generated using the in-house SZE tool octreemesher. All meshes use the same geometry, albeit at different resolutions, and are listed in Table \ref{tab:UAPMeshes}. All input data including mesh, weather boundary and pollution source are precalculated and present in files for the benchmark.

\begin{figure}[!htb]
\centering
\includegraphics[width=\columnwidth]{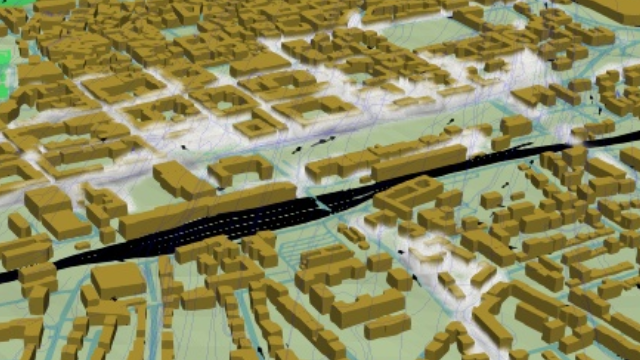}
\caption{Visualization of pollution dispersion in the Urban Air Project use case modeling the city of Győr. Streamlines and vectors indicate air flow direction, while white cloud shows pollution spread.}
\label{fig:uap3d}
\end{figure}

In this use case the RANS equations for incompressible fluids are solved with k-$\epsilon$ turbulence model. The solver simpleFoam is used, which implements the Semi-Implicit Method for Pressure Linked Equations (SIMPLE). Additionally, the RANS equations are coupled with convection-diffusion equations to model pollution spread. A total of $600$ iterations are run, except for high mesh size, where $400$ iterations are run, and runtime is adjusted afterwards. The Generalized Geometric-Algebraic Multigrid (GAMG) solver with Gauss-Seidel smoother is used to solve the pressure equations. A sample visualization of the results in 3D is shown on \ref{fig:uap3d}

\section{Experiments methodology}\label{section:Methodology}
In our experiments, all OpenFOAM kernels are compiled with the AOCC compiler and linked against the MPI library pre-installed by platform vendors. The AOCC compiler (v.4.1.0) is used with the optimization flag \texttt{-O3} and architecture-specific compiler arguments, including  \texttt{-march=znver2} for Rome, \\ \texttt{-march=znver3} for Milan and Milan~X, as well as \texttt{-march=znver4} for Genoa, Genoa~X and Bergamo. 

We investigate both application models (motorBike and UAP) on seven platforms by testing different problem sizes. The execution time is measured by extracting the time stamps written by OpenFOAM as “Execution Time” and subtracting the first value from the last value. This way time consumed by the initialization is discarded, albeit the runtime of one less iteration is measured. To ensure the reliability of measurements, every test is repeated 5 times and all measurements are reported and illustrated. 

All the systems operate with SMT disabled and turbo boost enabled. We observe that the thermal and power limitations of the test platforms allow them to operate at the frequency clock close to the maximum turbo boost speed (Table \ref{tab:cpuFreq}).

\begin{table}[!h]
\caption{Range of clock speed [GHz] measured during benchmark}
\label{tab:cpuFreq}
\begin{center}
\begin{tabular}{|c|c|}
\hline
Rome	       & 2.8 - 3.1 \\ \hline
Milan 	       & 2.9 - 3.2 \\ \hline
Milan~X	       & 2.8 - 3.0 \\ \hline
64-core Genoa  & $\approx$3.7 \\ \hline
64-core Genoa  & 3.4 - 3.5 \\ \hline
96-core Genoa  & 3.2 - 3.4 \\ \hline
128-core Genoa & 2.8 - 3.1 \\ \hline
\end{tabular}
\end{center}
\end{table}

To compare the performance between the various systems, the metric FVOPS (Finite VOlumes solved Per Second) is introduced \cite{Galeazzo2024}. The FVOPS metric is calculated as: $$FVOPS={{cell~count}\over{runtime~per~iteration~or~time~step}}$$ The value of FVOPS depends on a series of factors, including the simulation type, boundary conditions, and especially the grid size being solved. The results reported here use test cases with fixed conditions in all systems, only varying the grid size. To facilitate the comparison the amount of grid elements per core is the usual metric and is used in the X-axis. This metric often reveals local maxima, which indicates the optimal number of grid points per core per test case and architecture. It is interesting to note that these local maxima occur at different values of grid element per rank when utilizing different processor types.

In addition, we employ AMD $\mu$Prof (ver. 4.1) software profiling analysis tool \cite{uProf41} to reveal the applications' interaction with the hardware. The AMD $\mu$Prof performance tool analyses and monitors AMD Zen-based microarchitecture processors. In this work, we use a specific tool called AMDuProfPcm \cite{uProf41}, which allows fine-tuned monitoring of the CPU’s behaviours and identifies potential inefficiencies in AMD EPYC CPUs. This system analysis utility periodically collects the CPU, core, cache, and memory performance event count values and reports various metrics. 

In this stage of our work, we select a set of metric groups that includes: \textit{ipc}, \textit{l1}, \textit{l2}, \textit{l3}, \textit{tlb}, \textit{dc} and \textit{pipeline\_util}.  We distinguish metrics related to all L1, L2 and L3 levels of cache (access/hit/miss), core utilization, CPI (cycle per instruction), IPC (instruction per cycle), CPU pipeline, and others (see full specification in \cite{uProf41}). The selected groups of metrics enable us to indicate the impact of hardware features on performance efficiency.

\section{The motorBike use case benchmarking}

\subsection{Platforms comparison}

\begin{figure}[!htb]
\centering
\includegraphics[width=0.84\columnwidth]{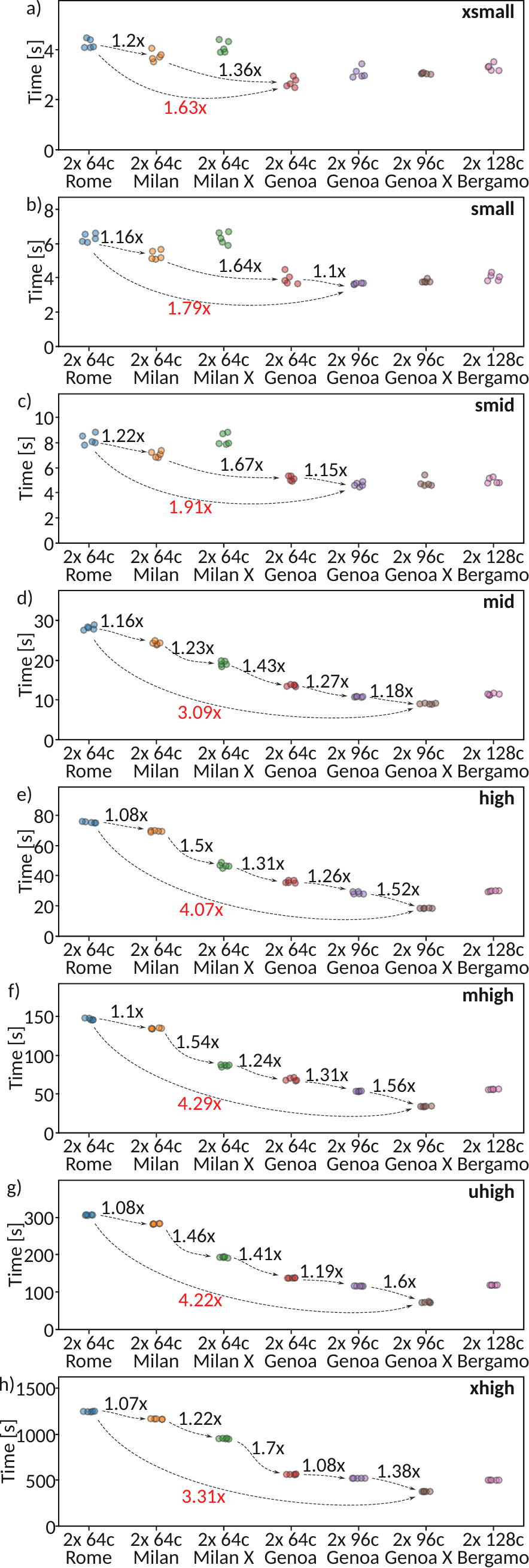}
\caption{Performance results obtained for motorBike on a variety of AMD CPUs and different mesh sizes}
\label{fig:motorTiming}
\end{figure}

Figure \ref{fig:motorTiming} presents the execution time measurements for the motorBike obtained for seven computing platforms (Table \ref{tab:platforms}) and eight mesh sizes (see Table \ref{tab:motorMeshes}). These results are grouped by the mesh size on different plots. Besides the execution time, this figure also reveals the performance comparison between the selected computing platforms. Notably, if noted, every plot outlines the performance gains for a given platform compared to the prior generation and the highest performance gain obtained between platforms with the best and the worst results.

Table \ref{tab:motorGains} delivers comprehensive performance comparisons across all platforms and different meshes. This table collates the range of speedups that indicate minimum and maximum performance differences for a fixed platform presented in the first row of the table compared to other systems outlined in the left column.


\begin{table*}
\caption{Aggregated performance gains of motorBike for different problem sizes and on a variety of AMD EPYC CPUs}
\label{tab:motorGains}
\begin{center}
\begin{tabular}{|c|c|c|c|c|c|c|c|}
\hline
	&	Rome	&	Milan	&	Milan~X	&	64-core Genoa	&	96-core Genoa	&	Genoa~X	&	Bergamo	\\ \hline
Rome	&	-	&	1.07x-1.22x	&	0.97x-1.69x	&	1.63x-2.22x	&	1.45x-2.75x	&	1.49x-4.29x	&	1.36x-2.62x	\\ \hline
Milan	&	0.82x-0.94x	&	-	&	0.81x-1.54x	&	1.35x-2.07x	&	1.21x-2.50x	&	1.24x-3.91x	&	1.14x-2.39x	\\ \hline
Milan~X	&	0.59x-1.04x	&	0.65x-1.24x	&	-	&	1.24x-1.70x	&	1.50x-1.92x	&	1.54x-2.67x	&	1.41x-1.91x	\\ \hline
64-core Genoa	&	0.45x-0.61x	&	0.48x-0.74x	&	0.59x-0.81x	&	-	&	0.89x-1.31x	&	0.91x-2.05x	&	0.84x-1.25x	\\ \hline
96-core Genoa	&	0.36x-0.69x	&	0.40x-0.83x	&	0.52x-0.67x	&	0.76x-1.13x	&	-	&	0.98x-1.60x	&	0.89x-1.04x	\\ \hline
Genoa~X	&	0.23x-0.67x	&	0.26x-0.81x	&	0.37x-0.65x	&	0.49x-1.10x	&	0.62x-1.02x	&	-	&	0.61x-0.92x	\\ \hline
Bergamo	&	0.38x-0.73x	&	0.42x-0.88x	&	0.52x-0.71x	&	0.80x-1.20x	&	0.96x-1.12x	&	1.09x-1.64x	&	-	\\ \hline
\end{tabular}
\end{center}
\end{table*}

Investigating the results, the platform with Rome CPUs exhibits the lowest performance for all mesh sizes except for small mesh sizes, where performance with Milan~X is tied (see Figure \ref{fig:motorTiming}b). The improvement of regular Milan over Rome is about a factor of up to 1.2x for smaller mesh sizes, which diminishes to 1.07x for large ones. Similar to Rome, Milan also uplifts negligible performance over Milan~X for the smaller meshes. The advantage of Milan~X compared to regular Milan and Rome is strongly noticeable for mid, high, mhigh, uhigh, and xhigh mesh sizes. In this case, the best performance profit equals 1.54x and 1.69x faster than Milan and Rome, respectively, and is observed for the mhigh mesh size. At the same time, the Milan~X processors do not improve performance analyzing the remaining smaller meshes.

Considering the AMD EPYC 9004 family (Genoa, Genoa~X and Bergamo), we observe significant performance improvement over all previous architectures across all mesh sizes. 
As shown in Table \ref{tab:motorGains}, the system with 64-core Genoa processors performs all the tests faster, accelerating computation about 1.63x-2.22x, 1.35x-2.07x, and 1.24x-1.70x than platforms with Rome, Milan, and Milan~X CPUs, respectively. We also identify that 64-core Genoa provides a performance advantage over the 96-core version, accelerating computation by about 1.3x faster for small mesh size. Considering other meshes in this comparison, the larger core count in 96-core Genoa brings a relatively negligible improvement compared to the 64-core version, improving performance in the range of 1.1x to 1.31x.

As shown in Table \ref{tab:motorGains} and Figure \ref{fig:motorTiming}, the platform with Genoa~X CPUs offers significant performance gain over other systems for mid, high, mhigh, high, and xhigh mesh sizes. This platform overcomes the prior generation, including Rome, Milan, and Milan~X CPUs, accelerating computation up to 4.29x, 3.91x, and 2.67x, respectively. This is also worth noting that -- similar to Milan~X -- the new Genoa~X model is not faster on smaller mesh sizes over regular Genoa processors. We also reveal that the 128-core Bergamo results are similar to those of the Genoa CPUs but with negligible performance drops over the 96-core model.

Table \ref{tab:motorBest} demonstrates the computing platforms that achieve the best results during tests performed for a given mesh size. As expected, the highest performance advantages are noticeable on platforms with Genoa~X CPUs for relatively larger mesh sizes. The regular Genoa CPUs, including 64- and 96-core models, bring the best performance improvements for the smaller ones.

\begin{table}[!h]
\caption{Platforms with the best results for motorBike}
\label{tab:motorBest}
\begin{center}
\begin{tabular}{|c|c|}
\hline
Platform & Meshes \\ \hline
2x 64-core AMD & \multirow{2}{*}{small} \\ 
EPYC 9554 (Genoa) & \\ \hline
2x 96-core AMD  & \multirow{2}{*}{small and smid}\\ 
EPYC 9654 (Genoa) & \\ \hline
2x 96-core AMD & mid, high, mhigh,  \\
EPYC 9684X (Genoa~X) & high, and xhigh \\ \hline
\end{tabular}
\end{center}
\end{table}

\subsection{Performance evaluation}

To compare performance across multiple platforms, we propose to use FVOPS performance metric. This metric helps us estimate computing efficiency for different mesh sizes, indicating better platform utilization for the higher FVOPS level.

\begin{figure}[!htb]
\centering
\includegraphics[width=0.8\columnwidth]{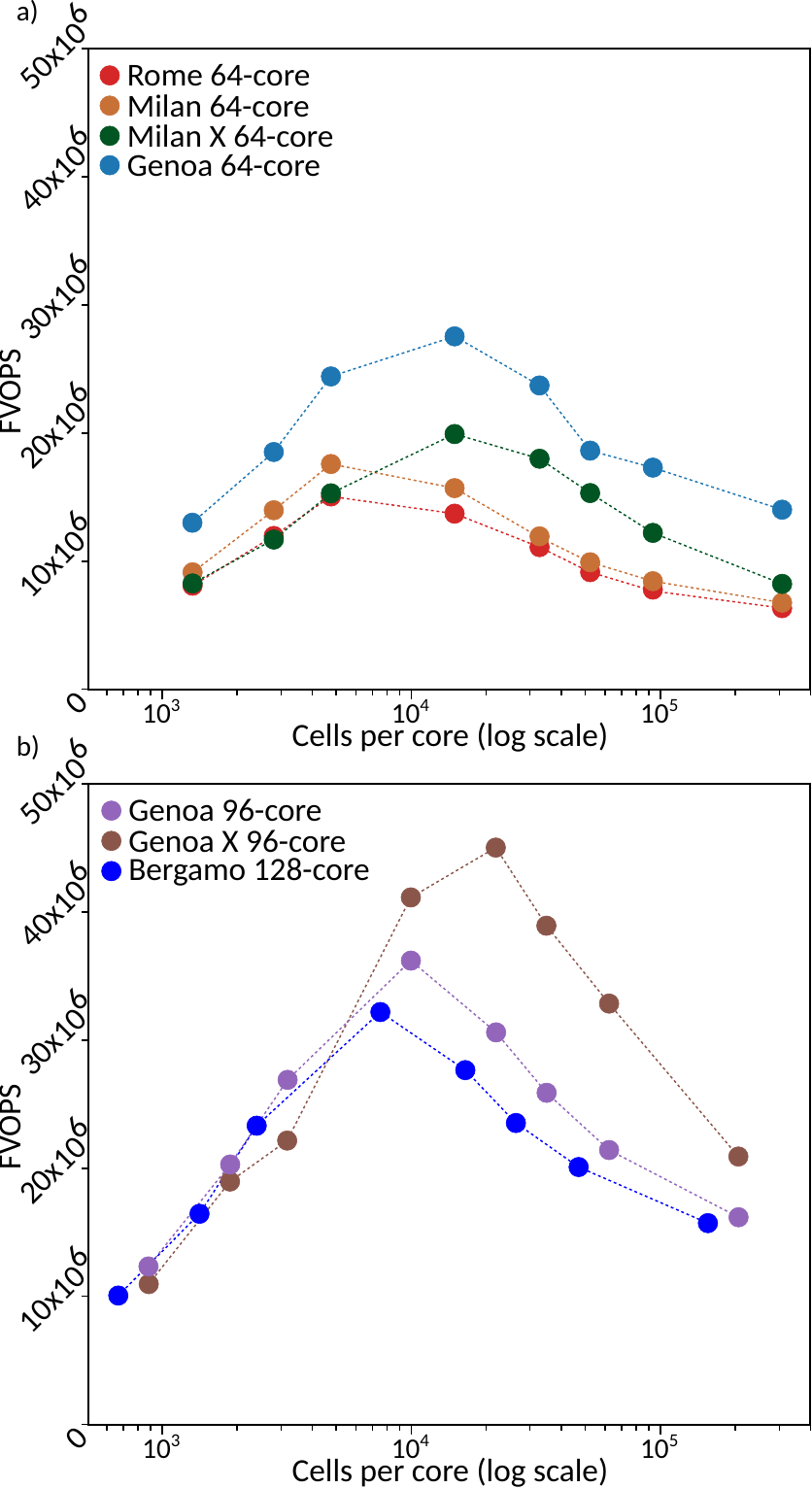}
\caption{FVOPS performance metric for motorBike}
\label{fig:motorFVOPS}
\end{figure}

Figure \ref{fig:motorFVOPS} reveals the calculated FVOPS for all platforms and mesh sizes. These results are grouped as 64-core architectures (Figure \ref{fig:motorFVOPS}a) and systems with 96-/128-cores CPUs (Figure \ref{fig:motorFVOPS}b). For every architecture, the calculation performance in FVOPS is plotted versus the per-core cell count.
 
FVOPS shows a similar trend for all architectures. It increases up to a turning point and decreases afterwards. The best platform utilization is noted for about 4715 cells per core (smid mesh) for Rome and Milan, 14821 (mid mesh) for Milan~X and 64-core Genoa, 7410 (mid mesh) for Bergamo, 9881 (mid mesh) for 96-core Genoa and 21700 (high mesh) for Genoa~X. For small cell count meshes, up to 5000, architectures before Genoa exhibit lower performance than 64-core Genoa. Differences between Rome, Milan and Milan~X are not that dominant. Furthermore, the 96-core Genoa and Genoa~X with Bergamo show similar performance at a given cell per core value, while outperforming 64-core Genoa. For larger mesh sizes, enabled 3D V-cache technology (Milan~X and Genoa~X) architectures dominate their own platform. Milan~X shows an advantage in the 15k to 100k cell per core range over regular Milan, while Genoa~X has an advantage in the 10k to 200k cell per core range over regular Genoa. This makes both products a viable choice for the cell count range of 2M to 13M and 1M to 20M, respectively. Bergamo shows no advantage on any of the investigated mesh sizes.

\begin{figure}[!htb]
\centering
\includegraphics[width=0.8\columnwidth]{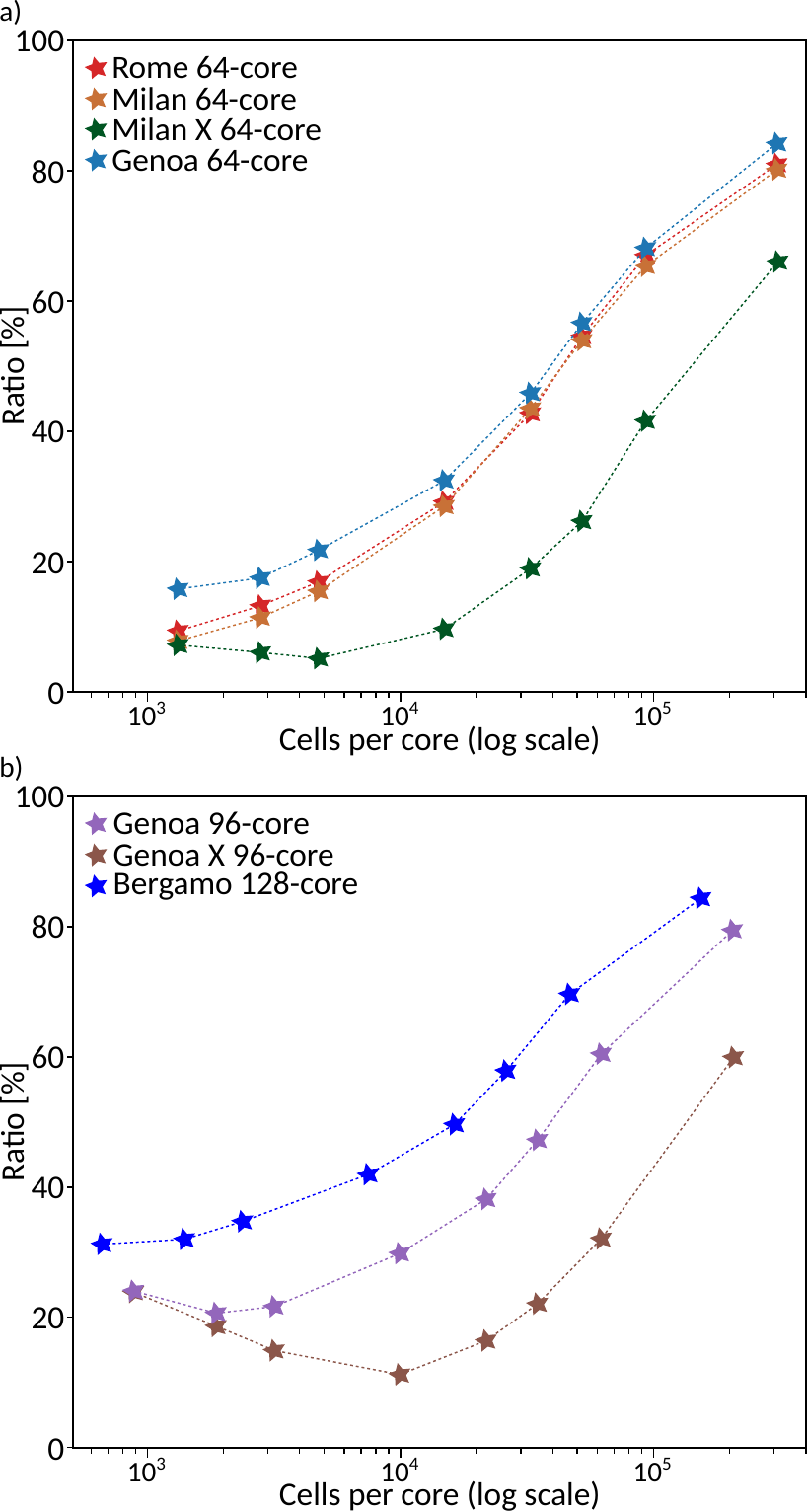}
\caption{L3 miss ratio for motorBike on a) systems with 64-core CPUs (Rome, Milan, Milan~X and Genoa) and b) platforms with 96-core and 128-core CPUs (Genoa, Genoa~X and Bergamo)}
\label{fig:motorL3}
\end{figure}

To explain the FVOPS results, a closer look at the hardware specification and performance evaluation is required. Hence, in this stage of our work, we start performance analysis of the impact of L3 cache capacity on performance. Figure \ref{fig:motorL3} illustrates the L3 cache miss ratio obtained for different meshes and a variety of computing systems, including platforms equipped with 64-core CPUs (Figure \ref{fig:motorL3}a) as well as 96- and 128-core CPUs (Figure \ref{fig:motorL3}b). Figure \ref{fig:motorL3}a shows that the L3 cache miss ratio follows a similar trend for systems with 64-core Rome, Milan, and Genoa CPUs. In this case, the L3 miss rate starts at around 10\%-15\% for the smaller mesh and consecutively grows to 80\% for the larger ones. These three platforms offer 256 MB of L3 for a single processor, thus featuring convergent L3 miss trend. In contrast, the Milan~X -- which offers 768 MB of L3 cache capacity per CPU -- reduces the L3 miss rate for larger mesh sizes. It is mainly noticeable for sizes mid, high, mhigh, and high, where up to 30 percentage points reduce the L3 miss rate over other 64-core CPUs. 

According to the FVOPS performance metric (Figure \ref{fig:motorFVOPS}), which decreases for relatively larger mesh sizes, the L3 cache size affects performance limits. This is because the total amount of data actively used by the software is greater than the size of the cache. Consequently, the overall performance is limited by the data traffic data volumes between L3 and main memory. Thus, both L3 cache size and the speed of the main memory subsystem play a key role in attainable performance. 

When comparing Milan~X with regular Milan and Rome, this results in a noticeable increase in performance mainly for larger mesh sizes (mid, high, mhigh, uhigh, and xhigh). However, such a performance advantage decreases when the data volume requirement exceeds the L3 capacity in Milan~X, reducing improvement over regular Milan from 1.54x for mhigh to 1.24x for xhigh mesh sizes. We should also note that the large L3 cache in Milan~X does not overcome regular Milan or Rome CPUs for relatively smaller sizes where the L3 cache capacity is not a critical performance bottleneck. Considering all studied platforms with 64-core CPUs (Figure \ref{fig:motorL3}a), the 4th generation of AMD EPYC processors, thanks to the newest DDR5 and larger 2x 12-channel memory subsystem, offers higher attainable performance than the prior generation Rome, Milan, and even Milan~X. 

Furthermore, considering rather smaller mesh sizes, we also observe that 96-core Genoa(X) and 128-core Bergamo processors do not offer a performance advantage over 64-core Genoa processors (see Figure \ref{fig:motorTiming}). Since the AMD EPYC 9004 series processors feature the same memory subsystem, this effect can be explained by the 64-core CPU offering more memory bandwidth per core than others to keep core-memory data movements more efficient.

Figure \ref{fig:motorL3}b demonstrates the L3 miss rate obtained for platforms with 96-core Genoa CPUs, 96-core Genoa~X CPUs and 128-core Bergamo CPUs. As expected, the best L3 cache miss ratio trend is notable for the platform with 3D V-cache technology (Genoa~X) compared with regular Genoa and Bergamo-based solutions. The large L3 capacity in Genoa~X CPUs features a smaller L3 cache miss rate of up to 28 and 37 percentage points over Genoa and Bergamo-based platforms, respectively. It results in noticeable performance profits over other platforms for sizes mid, high, mhigh, uhigh, and xhigh. Like Milan~X, the performance advantage of Genoa~X is not noticeable for smaller sizes where the L3 capacity is not the performance bottleneck.

It should be emphasized that the platform based in Bergamo exhibits the most unfavorable trend in L3 cache miss ratio among all platforms examined. This is mainly due to the smaller L3 cache size per core compared to other processors (see Table \ref{tab:cpus}). Considering the memory-bound nature of the motorBike, as expected, the L3 cache miss penalties significantly impact performance for the 2x 128 cores of the Bergamo-based platform that do not bring performance advantage over Genoa processors. We also observe that L3 misses occur mainly because the data volume required to transfer through the cache is larger than the total cache capacity. Additionally, for the platforms with Genoa~X processors and for rather smaller sizes, we expect that L3 inevitable misses are noticeable where the first time a memory location is read.

\begin{figure}[!htb]
\centering
\includegraphics[width=0.8\columnwidth]{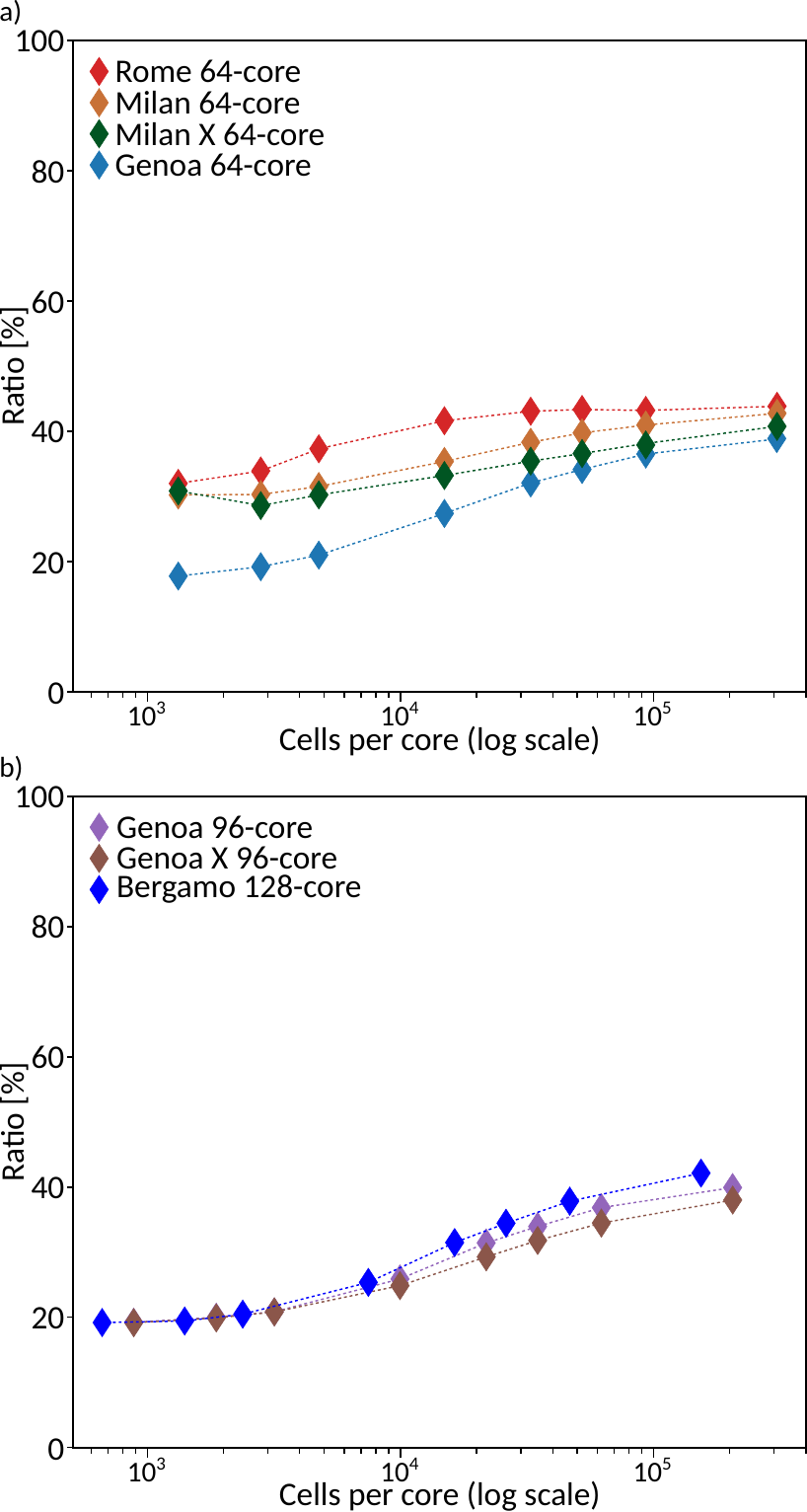}
\caption{Total L2 miss ratio for motorBike  on a) systems with 64-core CPUs (Rome, Milan, Milan~X and Genoa) and b) platforms with 96-core and 128-core CPUs (Genoa, Genoa~X and Bergamo)}
\label{fig:motorL2}
\end{figure}

The new AMD EPYC 9004 series processors have 1MB of L2 cache per core, which is twice as large as prior generations. As shown in Figure \ref{fig:motorL2}, it enables the reduction of the L2 cache miss rates and, hence, performance advantages compared to prior generations. In particular, the total L2 miss rate of Genoa and Bergamo CPUs is reduced by up to 12-14 percentage points over prior generation of CPUs for smaller mesh sizes (Figure \ref{fig:motorL2}a). Considering larger meshes, the total L2 miss trend becomes tied for all platforms with negligible advantage for Genoa-based processors. 


\section{Urban Air Pollution use case benchmarking}

\subsection{Platforms comparison}

\begin{figure}[!htb]
\centering
\includegraphics[width=0.84\columnwidth]{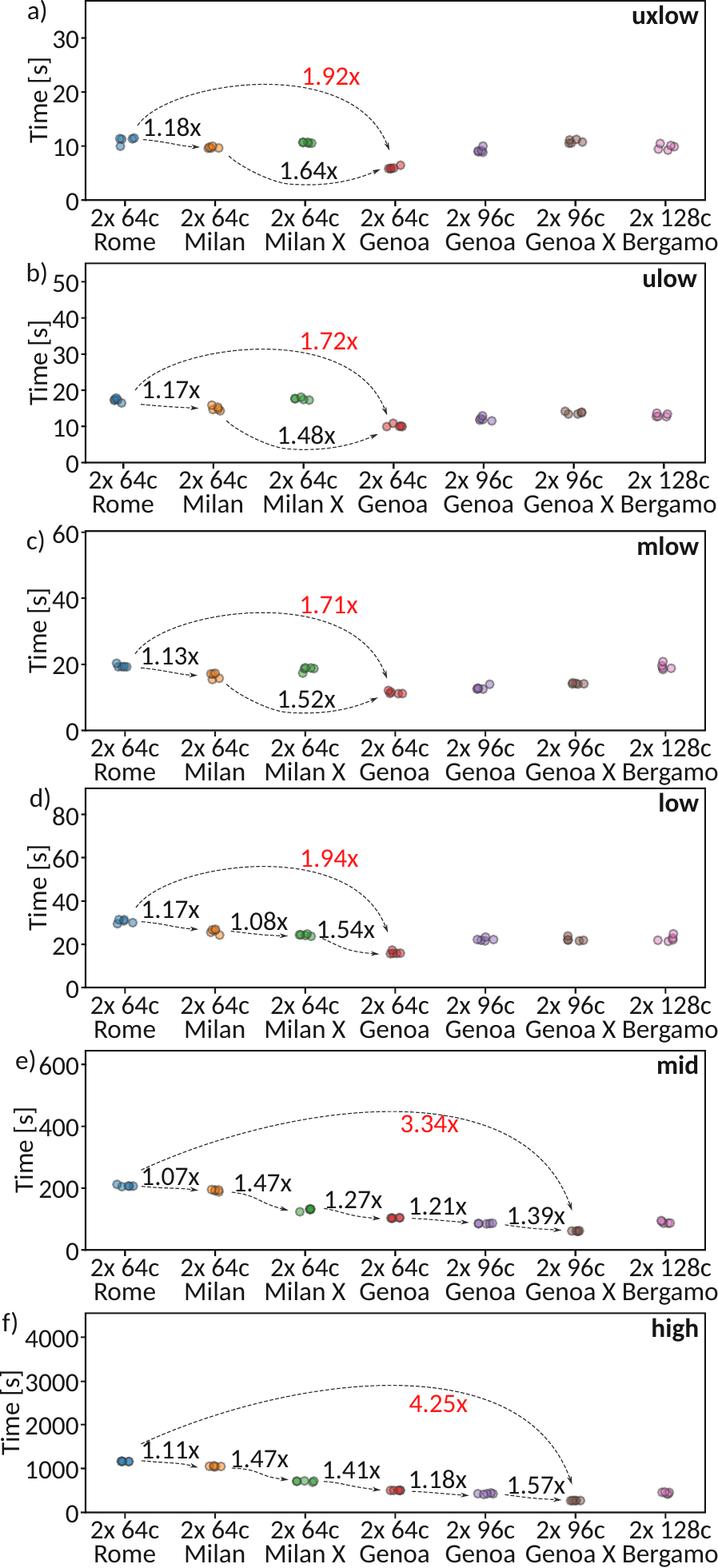}
\caption{Performance results obtained for UAP  on a variety of AMD CPUs and different mesh sizes}
\label{fig:uapTiming}
\end{figure}

The performance results obtained for the UAP on different computing platforms are presented in Figure \ref{fig:uapTiming}. This figure shows execution time measurements grouped by mesh size on various plots. When observed, we also demonstrate the performance improvements identified on a specific platform in comparison to the previous generation, as well as the greatest performance differences achieved across platforms with the best and worst results. Arrows indicate a larger than one speedup, including the actual speedup value.

Additionally, the comprehensive performance report is delivered in Table \ref{tab:uapGains}, indicating performance comparisons across all platforms and different meshes. This table collates the range of performance differences determined for a fixed platform presented in the first row of the table in comparison to other systems outlined in the left column.

As expected, the slowest execution time measurements are obtained for the platform with Rome CPUs. The regular Milan CPUs accelerate computation over Rome by around 1.07x-1.18x faster. For smaller meshes up to mlow size, we do not note performance profits for the system with Milan~X CPUs over other platforms. In contrast, a noticeable improvement is observed for larger meshes including low, midm, and high, with a speedup factor of up to 1.47x and 1.63x over regular Milan and Rome respectively.

As shown in Figure \ref{fig:uapTiming} and Table \ref{tab:uapGains}, the 64-core Genoa delivers the best performance for smaller meshes up to the low size over both the prior generation of CPUs as well as 96-core Genoa, Genoa~X, and Bergamo CPUs. The highest performance leap is observed for uxlow size, where the 64-core Genoa CPUS accelerates computation with a speedup factor of 2.30x, 2.08x, and 1.80x faster over Rome, Milan, and Mila X. In this case, we also indicate performance gains of about 1.54x, 1.82x, and 1.69x faster than 96-core Genoa, Genoa~X, and Bergamo CPUs, respectively. We also reveal that 96-core Genoa and Genoa~X do not provide the extra performance seen for smaller meshes, and usually perform worse or similar results in comparison to 64-core CPUs. 

In contrast, considering mid and high mesh sizes, the 96-core Genoa provides a circa 20\% benefit over the 64-core Genoa. Furthermore, the high cache 96-core Genoa~X shows similar behavior with a performance improvement of 39\% and 57\% over the 96-core Genoa for mid and high mesh sizes. In this case, Genoa~X also features noticeable performance leaps with a speedup factor of up to 1.85x than 64-core Genoa CPUs. Finally, at neither of the mesh sizes does Bergamo deliver the best performance.

\begin{table*}
\caption{Aggregated performance gains of UAP for different problem sizes and on a variety of AMD EPYC CPUs}
\label{tab:uapGains}
\begin{center}
\begin{tabular}{|c|c|c|c|c|c|c|c|}
\hline
	&	Rome	&	Milan	&	Milan~X	&	64-core Genoa	&	96-core Genoa	&	Genoa~X	&	Bergamo	\\ \hline
Rome    &-            & 1.07x-1.18x & 0.98x-1.63x & 1.72x-2.30x & 1.24x-2.71x & 1.05x-4.25x & 1.02x-2.54x \\ \hline 
Milan   & 0.85x-0.94x &-            & 0.84x-1.47x & 1.48x-2.08x & 1.06x-2.44x & 0.90x-3.83x & 0.90x-2.29x \\ \hline
Milan~X & 0.61x-1.02x & 0.68x-1.19x &-            & 1.27x-1.80x & 1.10x-1.66x & 0.99x-2.61x & 0.99x-1.56x \\ \hline
64-core Genoa   & 0.43x-0.58x & 0.48x-0.68x & 0.55x-0.79x &-            & 0.65x-1.21x & 0.55x-1.85x & 0.59x-1.18x \\ \hline
96-core Genoa   & 0.37x-0.80x & 0.41x-0.94x & 0.60x-0.91x & 0.83x-1.54x &-            & 0.85x-1.57x & 0.67x-1.00x \\ \hline
Genoa~X & 0.24x-0.95x & 0.26x-1.11x & 0.38x-1.01x & 0.54x-1.82x & 0.64x-1.18x &-            & 0.60x-1.09x \\ \hline
Bergamo & 0.39x-0.98x & 0.44x-1.11x & 0.64x-1.01x & 0.85x-1.69x & 1.00x-1.49x & 0.92x-1.67x &-\\ \hline
\end{tabular}
\end{center}
\end{table*}

The list of the best computing platforms that feature the best results for a given mesh size is outlined in Table \ref{tab:uapBest}. We reveal that the platform with 64-Genoa CPUs offers the best results when processing relatively small problem sizes. The large L3 capacity of Genoa~X CPUs brings the highest performance advantages for larger mesh sizes.

\begin{table}[!h]
\caption{Platforms with the best results for motorBike}
\label{tab:uapBest}
\begin{center}
\begin{tabular}{|c|c|}
\hline
Platform & Meshes \\ \hline
2x 64-core AMD & \multirow{2}{*}{uxlow, ulow, mlow, and low} \\ 
EPYC 9554 (Genoa) & \\ \hline
2x 96-core AMD & \multirow{2}{*}{mid and high}   \\
EPYC 9684X (Genoa~X) &  \\ \hline
\end{tabular}
\end{center}
\end{table}

\subsection{Performance evaluation}

\begin{figure}[!htb]
\centering
\includegraphics[width=0.8\columnwidth]{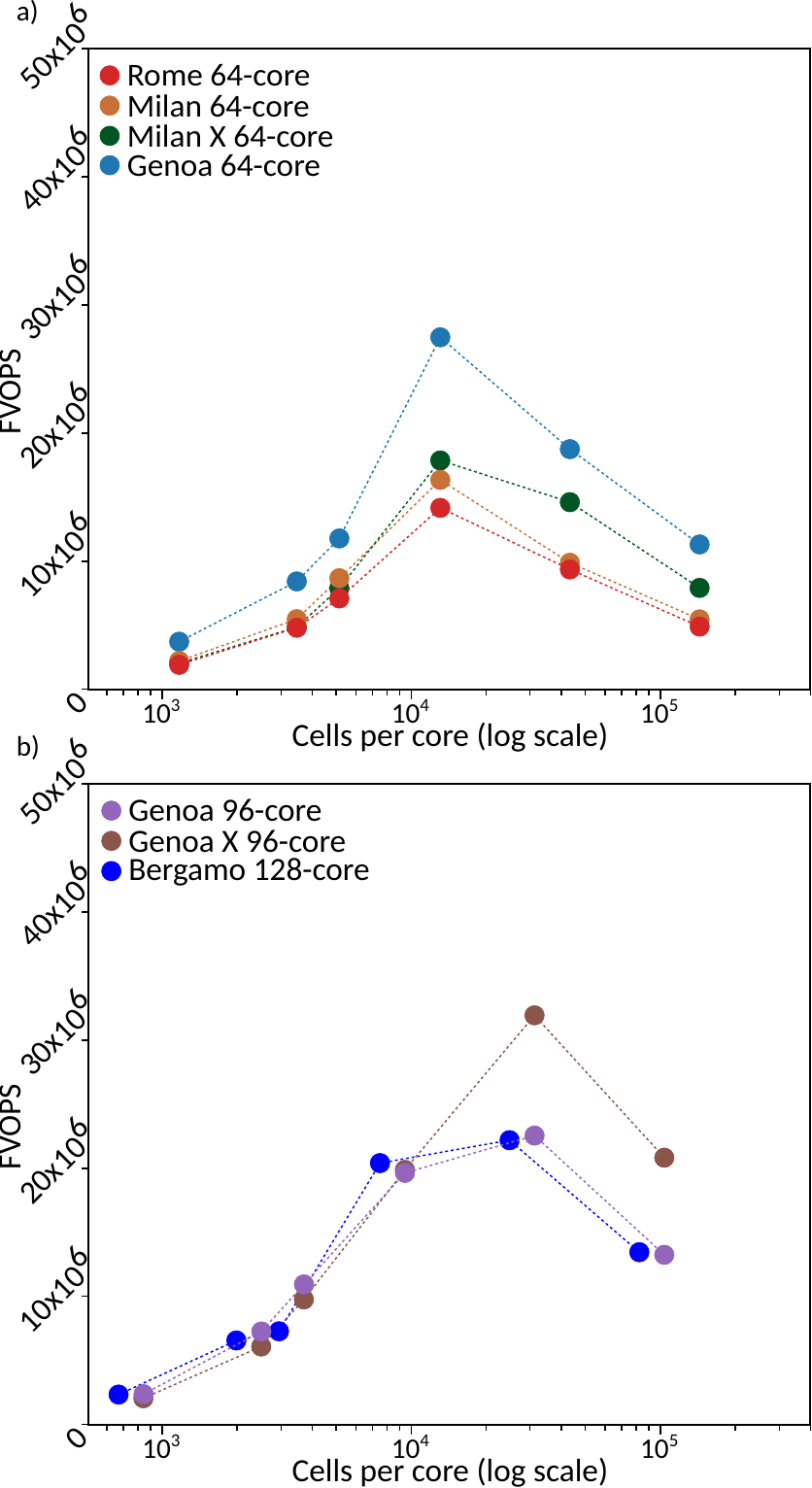}
\caption{FVOPS performance metric for UAP}
\label{fig:uapFVOPS}
\end{figure}

As with the motorBike benchmark, we use the FVOPS metric to indicate computing efficiency for different mesh sizes across multiple platforms. Figure \ref{fig:uapFVOPS} shows FVOPS for the UAP benchmark. These results are grouped again as 64-core architectures (\ref{fig:uapFVOPS}a) and 96-/128-cores CPUs (\ref{fig:uapFVOPS}b), where FVOPS values are plotted against per core cell count (see Table \ref{tab:UAPMeshes}).

As shown in Figure \ref{fig:uapFVOPS}, the FVOPS curves show similar behaviour for all platforms, having a maximum value between ca. 6k and 20k cells per core, and decreasing for lower and higher core cell counts. More precisely, for all 64-core CPUs, the peak parallel efficiency is observed at 5688 cells per core (Figure \ref{fig:uapFVOPS}a), which corresponds to the low mesh size. In this case, the low mesh size features a higher utilization of 64-core platforms regarding the number of processed cells per second by a single core. Moreover, the Genoa architecture dominates among the 64-core results, while Milan~X does have an advantage over Rome and Milan for the largest two meshes. For smaller cell per core counts, Rome and Milan do not exhibit significant differences.

The results for processors with higher core counts are almost identical, except for the Genoa~X processor, which outperforms other processors with the largest mesh sizes. The maximum FVOPS is noted for the mid mesh size and refers to 16808 and 12606 core cell counts for platforms with 96-core and 128-core CPUs, respectively.
These platforms process the mid mesh with higher efficiency - in terms of the number of processed cells per second by core - compared to other meshes. 
In addition, it can be observed that providing more cores for calculations does not bring performance improvement as 96-core Genoa and 128-core Bergamo cannot show higher performance within this metric. 

\begin{figure}[!htb]
\centering
\includegraphics[width=0.8\columnwidth]{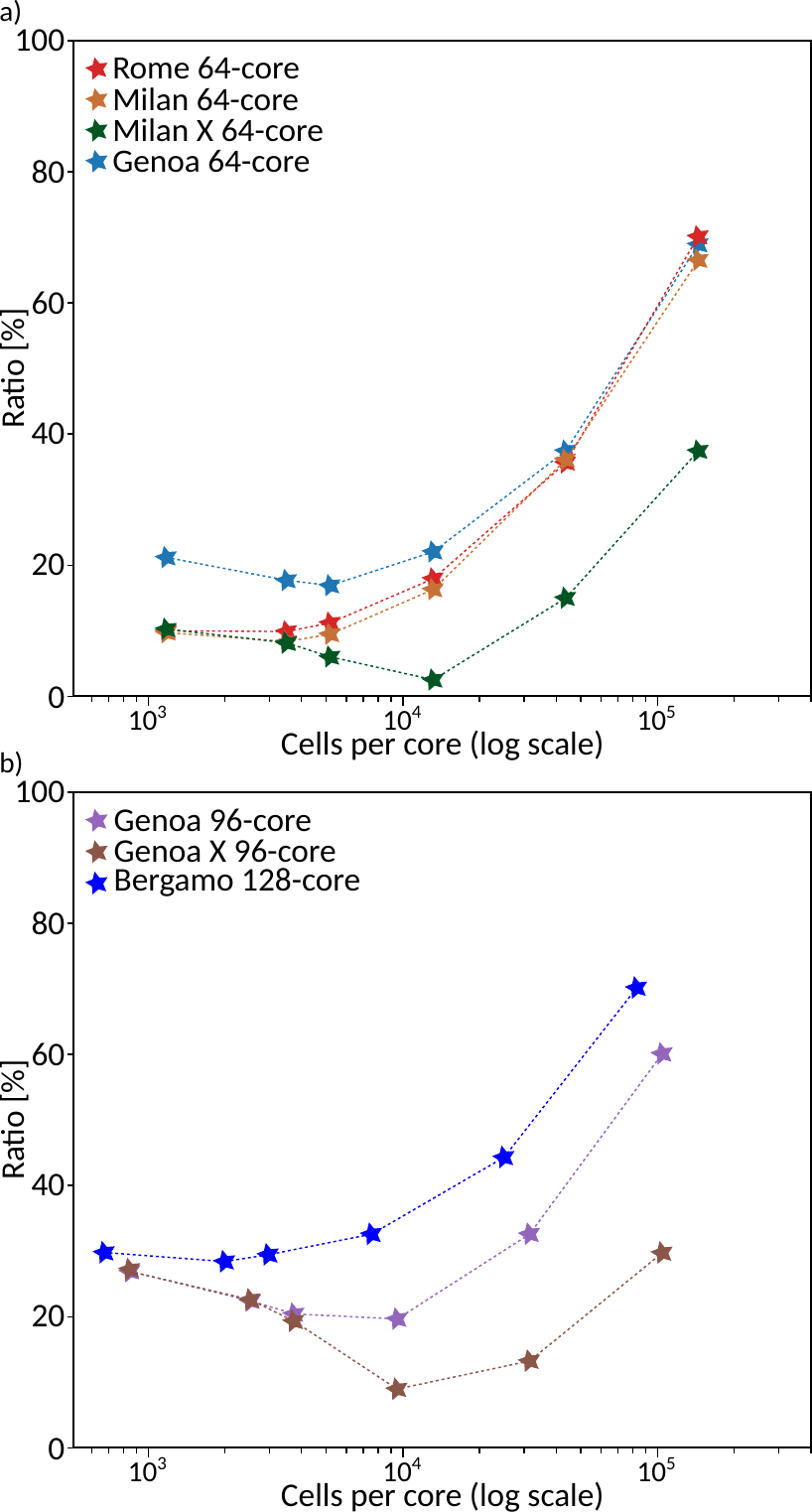}
\caption{L3 miss ratio for UAP on a) systems with 64-core CPUs (Rome, Milan, Milan~X and Genoa) and b) platforms with 96-core and 128-core CPUs (Genoa, Genoa~X and Bergamo)}
\label{fig:uapL3}
\end{figure}

Considering relatively large domain sizes, including mid and high, we observe a high L3 cache miss rate that reaches up to 70\% (Figure \ref{fig:uapL3}). In this case, following the trend of FVOPS performance metrics, the capacity of the L3 cache plays a key role in the attainable performance. This is because the application requirement for the data volume exceeds cache capacity and generates mainly L3 capacity misses. In contrast, considering relatively smaller domain sizes, we observe that the L3 miss ratio is kept at a smaller rate. We expect here that L3 misses are mainly noticeable where the first time a memory location is read (compulsory misses). As a result, the traffic through the L3 and main memory strongly affects overall performance across all sizes.

However, this hardware constraint is alleviated by enabling 3D V-cache technology in Milan~X and Genoa~X processors. In this case, we note a reduction in the L3 cache miss rate by up to 29 and 30 percentage points over regular Milan and Genoa CPUs, respectively. This results in a performance advantage by reducing the cost of data movement and accelerating the computation of up to 1.47x and 1.56x, respectively, for Milan~X and Genoa~X compared to Milan and Genoa CPUs.

Consequently, the novel AMD EPYC 9004 series processors thanks to the newest DDR5-based and larger 2x 12-channel memory subsystem offer higher attainable performance than the prior generation (Rome, Milan, and Milan~X). We note the performance uplift for the 96-core EPYC Genoa CPUs of up to 2.5x and 2.7x over regular Milan and Rome CPUs, mainly resulting from the 12-channel DDR5 and an extra 32 cores.

As expected, the Bergamo-based platform features the highest trend for the L3 cache miss ratio (Figure \ref{fig:uapL3}). To explain this behaviour, we have to look at the specification of the Bergamo architecture that offers a smaller L3 cache size per core and the same memory subsystem speed compared to other Genoa-based processors (see Table \ref{tab:cpus}). Taking this into account, and the fact that memory-intensive parallel codes can suffer from bandwidth saturation as more cores are used, Bergamo processors do not provide a performance advantage over Genoa processors.

\begin{figure}[!htb]
\centering
\includegraphics[width=0.8\columnwidth]{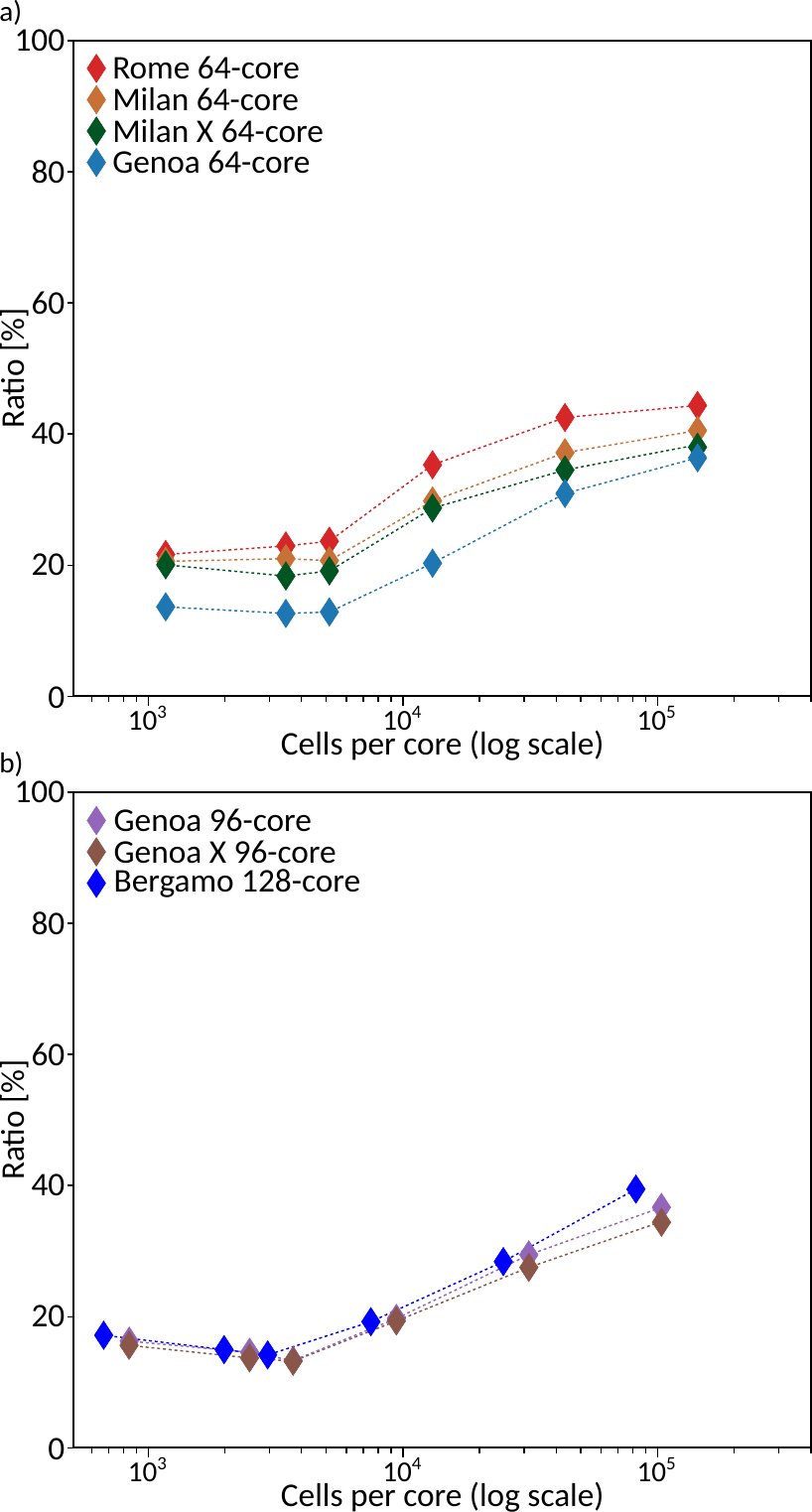}
\caption{Total L2 miss ratio for UAP on a) systems with 64-core CPUs (Rome, Milan, Milan~X and Genoa) and b) platforms with 96-core and 128-core CPUs (Genoa, Genoa~X and Bergamo)}
\label{fig:uapL2}
\end{figure}

The Genoa-based processors offer twice larger L2 size per core, reducing L2 cache miss ratio trends compared to the prior AMD EPYC generations (Figure \ref{fig:uapL2}). This trend reduction is observed for all the studied metrics, including the L2 miss ratio from L1DC, the L2 miss ratio from HWPF, and the L2 miss ratio total. It results in performance advantages over the prior generations of AMD EPYC CPUs, especially noted for relatively smaller mesh sizes (see Figure \ref{fig:uapTiming}).

\section{Cross-platform analysis of metrics}

This section aims to assess the impact of a wide range of performance metrics provided by the AMD profiler on the FVOPS metric introduced in Section~\ref{section:Methodology}. To gain a better understanding of FVOPS behaviour, we focus on two main objectives. Firstly, we analyze the correlation of individual metrics with FVOPS to identify metrics with the greatest influence on overall FVOPS behaviour. Secondly, we delve deeper into the issue by concentrating on two complementary questions: which metrics have the most decisive impact on the growth of FVOPS, and analogously which metrics are most influential in decreasing FVOPS. Addressing these objectives serves as starting point towards developing the FVOPS prediction model.

To this end, we analyze the metrics collected using the AMD $\mu$Prof software profiling tool for both motorBike and UAP use cases across all examined computing platforms presented in Table~\ref{tab:cpus}. The number of available performance metrics varies slightly between different computing platforms, but generally includes about fifty different metrics, with a particular focus on those providing details related to various levels of cache memory. 

For evaluating the correlation between individual metrics and FVOPS, we utilize the Pearson correlation coefficient \cite{Pearson}. The values of this coefficient range from -1 to 1, where -1 indicates a perfect negative linear relationship, 0 indicates no linear relationship, and 1 indicates a perfect positive linear relationship between the two variables being compared. Tables~\ref{tab:motorbike_Pearson} and~\ref{tab:UAP_Pearson} summarize the analysis, presenting the Pearson coefficient values for three metrics that perform best in both use cases. As can be observed, each metric refers to a different cache level. Starting from the L1 cache, we observe that the correlation coefficient for the metric, which determines the total number of DC fills from the same CCX, achieves an average of 0.79 for motorBike and 0.87 for UAP, indicating the significant impact of this metric on shaping FVOPS. Similar values (0.80 and 0.86, respectively) are recorded for the metric concerning the number of L2 hits caused by missing data in the L1 DC level. In both use cases, the best results in terms of correlation are achieved by the L3 hit metric, with an average of 0.86 for motorBike and 0.94 for UAP.

\begin{table}[!h]
\caption{Pearson coefficients between FVOPS and presented metrics for motorBike on different computing platforms}
\label{tab:motorbike_Pearson}
\begin{center}
\begin{tabular}{|c|c|c|c|}
\hline
\multirow{3}{*}{Platform} & DC Fills    & L2 hit    & \multirow{3}{*}{L3 hit}   \\ 
                          & from        & from      &                           \\
                          & same CCX    & DC miss   &                           \\\hline     
Rome            & --    & 0.89  & 0.80 \\ \hline
Milan           & 0.81  & 0.80  & 0.85 \\ \hline
Milan~X         & 0.69  & 0.74  & 0.91 \\ \hline
64-core Genoa   & 0.88  & 0.83  & 0.84 \\ \hline
96-core Genoa   & 0.71  & 0.69  & 0.85 \\ \hline
Genoa~X         & 0.88  & 0.88  & 0.97 \\ \hline
Bergamo         & 0.79  & 0.77  & 0.82 \\ \hline \hline
Average         & 0.79  & 0.80  & 0.86 \\ \hline
\end{tabular}
\end{center}
\end{table}

\begin{table}[!h]
\caption{Pearson coefficients between FVOPS and presented metrics for UAP on different computing platforms}
\label{tab:UAP_Pearson}
\begin{center}
\begin{tabular}{|c|c|c|c|}
\hline
\multirow{3}{*}{Platform} & DC Fills    & L2 hit    & \multirow{3}{*}{L3 hit}   \\ 
                          & from        & from      &                           \\
                          & same CCX    & DC miss   &                           \\\hline   
Rome            & --    & 0.86  & 0.85 \\ \hline
Milan           & 0.92  & 0.90  & 0.89 \\ \hline
Milan~X         & 0.97  & 0.97  & 0.99 \\ \hline
64-core Genoa   & 0.83  & 0.82  & 0.96 \\ \hline
96-core Genoa   & 0.79  & 0.78  & 0.96 \\ \hline
Genoa~X         & 0.88  & 0.88  & 0.96 \\ \hline
Bergamo         & 0.80  & 0.81  & 0.94 \\ \hline \hline
Average         & 0.87  & 0.86  & 0.94 \\ \hline
\end{tabular}
\end{center}
\end{table}

To address the second objective, which is to identify metrics that have a dominant impact on the increase and decrease in FVOPS, we investigate the changes in individual performance metrics accompanying the changes in FVOPS. If we look at the behaviour of the FVOPS (presented in Figure~\ref{fig:motorFVOPS} for motorBike and Figure~\ref{fig:uapFVOPS} for UAP) we observe that this metric follows the same typical pattern for all architectures. Initially, it rises with an increase in cells per core, peaks and then begins to decline. For this reason, we divide the analysis into two independent stages. First, for each architecture, we determine the mesh size at which we record the best FVOPS value. This mesh size serves as a reference point from which we create two groups of data: \textsc{L}, containing all data for smaller mesh sizes, and \textsc{R}, including all data for bigger mesh sizes. Then, within each group, we look for the mesh size with the lowest FVOPS value and compare the data concerning this point with the data related to the reference point. This enables us to identify metrics that change the most within each group. 

Tables~\ref{tab:left} and~\ref{tab:right} provide a summary of this analysis for the \textsc{L} and \textsc{R} groups, respectively. Each number in the tables represents the change in a given metric, calculated as an average across all computing platforms. Values greater than 1 indicate an increase, while smaller values indicate a decrease with respect to the reference point.

\begin{table}[!h]
\caption{Metrics that change the most with a increase of FVOPS (group L)}
\label{tab:left}
\begin{tabular}{|c|c|c|c|}
\hline
Category            & Metric                                                           & motorBike    & UAP          \\ \hline
--                  & FVOPS                                                            & 2.70$\times$ & 9.21$\times$ \\ \hline
\multirow{2}{*}{L1} & All DC Fills                                                     & 2.01$\times$ & 2.91$\times$ \\ \cline{2-4} 
                    & \begin{tabular}[c]{@{}c@{}}DC Fills\\ from same CCX\end{tabular} & 1.93$\times$ & 2.79$\times$ \\ \hline
\multirow{6}{*}{L2} & Access                                                           & 2.13$\times$ & 3.06$\times$ \\ \cline{2-4} 
                    & Access from HWPF                                                 & 2.88$\times$ & 6.92$\times$ \\ \cline{2-4} 
                    & Access from DC Miss                                              & 1.93$\times$ & 2.85$\times$ \\ \cline{2-4} 
                    & Hit                                                              & 1.89$\times$ & 2.58$\times$ \\ \cline{2-4} 
                    & Hit from HWPF                                                    & 2.27$\times$ & 5.06$\times$ \\ \cline{2-4} 
                    & Hit from DC Miss                                                 & 1.88$\times$ & 2.75$\times$ \\ \hline
\multirow{2}{*}{L3} & Access                                                           & 2.28$\times$ & 4.17$\times$ \\ \cline{2-4} 
                    & Hit                                                              & 2.04$\times$ & 4.02$\times$ \\ \hline
\end{tabular}
\end{table}

\begin{table}[!h]
\caption{Metrics that change the most with a decrease of FVOPS (group R)}
\label{tab:right}
\begin{tabular}{|c|c|c|c|}
\hline
Category               & Metric                                                                                 & motorBike             & UAP                    \\ \hline
--                     & FVOPS                                                                                  & 0.44$\times$                  & 0.48$\times$                   \\ \hline
\multirow{2}{*}{L1}    & \multirow{2}{*}{\begin{tabular}[c]{@{}c@{}}DC Fills \\ From Local Memory\end{tabular}} & \multirow{2}{*}{9.92$\times$} & \multirow{2}{*}{13.66$\times$} \\
                       &                                                                                        &                       &                        \\ \hline
\multirow{3}{*}{L2}    & Miss                                                                                   & 1.58$\times$                  & 1.28$\times$                   \\ \cline{2-4} 
                       & Miss from HWPF                                                                         & 1.60$\times$                   & 1.32$\times$                   \\ \cline{2-4} 
                       & Miss from DC Miss                                                                      & 2.13$\times$                  & 1.60$\times$                    \\ \hline
L3                     & Miss                                                                                   & 2.12$\times$                  & 3.33$\times$                   \\ \hline
\multirow{2}{*}{Other} & CPI                                                                                    & 3.11$\times$                   & 1.71$\times$                    \\ \cline{2-4} 
                       & \begin{tabular}[c]{@{}c@{}}Backend\_Bound - \\ .Memory\end{tabular}                    & 2.06$\times$                   & 1.71$\times$                    \\ \hline
\end{tabular}
\end{table}

Starting from the Table~\ref{tab:left}, we observe that in both use cases, the increase in FVOPS is accompanied by an increase in metrics related to the Fills of L1 DC, as well as Accesses and Hits of L2 and L3 cache levels. For instance, with an average increase in FVOPS by 2.70x in the motorBike use case, we also record an average increase in metrics: All DC Fills by 2.01x, L2 Access by 2.13x, and L3 Access by 2.28x.

Moving on to the Table~\ref{tab:right}, which summarizes the analysis of group \textsc{R}, we observe an average decrease in FVOPS to 0.44x and 0.48x of the best recorded value for motorBike and UAP, respectively. Within this group, we identify 7 other metrics that exhibit significant changes. In the L1 category, we observe a notable increase in the 'DC Fills from Local Memory' metric (9.92x and 13.66x), attributed to exceeding the capacities of L3 memory as the size of the analyzed mesh increases. The next category concerns other metrics that are not directly related to individual cache levels. The first of these is CPI, which describes the total number of cycles divided by the number of instructions. Low values of CPI indicate better computational performance. We note that the aforementioned decreases in FVOPS are accompanied by increases in the CPI metric by an average of 3.11x for motorBike and 1.71x for UAP. Additionally, we observe increases for both use cases (2.06x and 1.71x) in the Backend\_Bound.Memory metric, which determines the fraction of dispatched slots that remained unused due to stalls in the memory subsystem. The behaviour of this metric confirms the significant impact of memory efficiency on both applications.

\section{Conclusions}

The work presents an empirical analysis of the performance of CFD applications based on two selected models (motorBike and Urban Air Pollution) for different generations of AMD processor architectures of the professional EPYC line. Particular attention was paid to the impact of cache (L2 and L3), RAM (DDR4 and DDR5) and the related bandwidth of memory channels, which is important for memory-bound applications.
First, the impact of the FVOPS metric was assessed with changing problem size. It was noticed that the application performance increases with decreasing grid size up to a certain point resulting from the saturation of both the cache memory (specifically L3) and the bandwidth of the memory channels. After reaching the saturation (inflection) point (individual for different processors) resulting from the amount of data per one computational unit (core), a decrease in processing efficiency was recorded. It should be assumed that this is the result of limiting the available memory bandwidth per processing unit and the related deterioration of the effectiveness of the prediction algorithm for the L3 cache, which resulted in higher L3 miss rates (up to 80\%).
The research conducted by means of the \textmu Prof profiling application confirmed the assumption that a larger cache is beneficial for achieving higher performance. However, it should be noted that increasing the number of cores and cache size in the CPU does not result in proportionally higher performance, which is due to limitations in data flow and a smaller amount of L3 cache per core. This was particularly visible when measuring the L3 misses metric, which was related to the fact that the data volume required to transfer through the cache is larger than the total cache capacity. 

We also reveal that the traffic through the L3 and main memory strongly affects overall performance across all sizes. According to FVOPS, we identify the mesh size with the highest computing efficiency. Considering all platforms, we observe the average decrease in FVOPS to 0.44x and 0.48x of the best-recorded value for motorBike and UAP, respectively. Such decreases in computational performance can be explained by the increasing demand for significant amounts of data, given the larger mesh sizes that exceed cache capacity at higher levels.

The cross-platform analysis of memory-related metrics shows that, depending on the size of the tested task, different groups of factors have a significant impact on the achieved performance regardless of the CPU model. For smaller mesh sizes, metrics related to successful access to data ("access" and "hit" metrics) stand out, while for larger sizes, those related to failed data retrieval ("miss" metrics) from the L2 and L3 cache become more important. In both cases, the metric related to filling the data cache (DC cache Fills) played a significant role.
Contrary to expectations, the latest AMD architecture (EPYC 9754 - Bergamo) did not prove to be the most efficient for both smaller and larger mesh sizes. The best performance was measured with the Genoa architecture - EPYC 9554 for small mesh sizes and EPYC 9684X for larger ones.

\section*{Acknowledgements}

Funded by the European Union. This work has received funding from the European High Performance Computing Joint Undertaking and Poland, Germany, Spain, Hungary, France and Greece under grant agreement number: 101093457. This publication expresses the opinions of the authors and not necessarily those of the EuroHPC JU and Associated Countries which are not responsible for any use of the information contained in this publication.
The authors are grateful to AMD company for granting access to the HPC platforms.

\bibliographystyle{elsarticle-num}
\bibliography{references.bib}

\begin{thebibliography}{10}
\expandafter\ifx\csname url\endcsname\relax
  \def\url#1{\texttt{#1}}\fi
\expandafter\ifx\csname urlprefix\endcsname\relax\def\urlprefix{URL }\fi
\expandafter\ifx\csname href\endcsname\relax
  \def\href#1#2{#2} \def\path#1{#1}\fi

\bibitem{TOP500}
\href{https://top500.org/}{{TOP500 website}}, {Accessed}: 2024-05-17 (2024).
\newline\urlprefix\url{https://top500.org/}

\bibitem{EuroHPC}
\href{https://eurohpc-ju.europa.eu/}{{EuroHPC Joint Undertaking website}},
  {Accessed}: 2024-05-17 (2024).
\newline\urlprefix\url{https://eurohpc-ju.europa.eu/}

\bibitem{OpenFOAMwebsite}
\href{https://www.openfoam.com/}{{OpenFOAM website}}, {Accessed}: 2024-05-17
  (2024).
\newline\urlprefix\url{https://www.openfoam.com/}

\bibitem{MARTINS2014218}
N.~M. Martins, N.~J. Carriço, H.~M. Ramos, D.~I. Covas,
  \href{https://www.sciencedirect.com/science/article/pii/S0045793014003764}{Velocity-distribution
  in pressurized pipe flow using cfd: Accuracy and mesh analysis}, Computers
  and Fluids 105 (2014) 218--230.
\newblock \href
  {http://dx.doi.org/https://doi.org/10.1016/j.compfluid.2014.09.031}
  {\path{doi:https://doi.org/10.1016/j.compfluid.2014.09.031}}.
\newline\urlprefix\url{https://www.sciencedirect.com/science/article/pii/S0045793014003764}

\bibitem{Moureau2011}
V.~Moureau, P.~Domingo, L.~Vervisch,
  \href{https://comptes-rendus.academie-sciences.fr/mecanique/articles/10.1016/j.crme.2010.12.001/}{Design
  of a massively parallel cfd code for complex geometries}, Comptes Rendus.
  Mécanique 339 (2011) 141--148.
\newblock \href {http://dx.doi.org/https://10.1016/j.crme.2010.12.001}
  {\path{doi:https://10.1016/j.crme.2010.12.001}}.
\newline\urlprefix\url{https://comptes-rendus.academie-sciences.fr/mecanique/articles/10.1016/j.crme.2010.12.001/}

\bibitem{Hadade2020}
I.~Hadade, T.~M. Jones, F.~Wang, L.~d. Mare,
  \href{https://doi.org/10.1145/3380932}{Software prefetching for unstructured
  mesh applications}, ACM Trans. Parallel Comput. 7~(1).
\newblock \href {http://dx.doi.org/10.1145/3380932}
  {\path{doi:10.1145/3380932}}.
\newline\urlprefix\url{https://doi.org/10.1145/3380932}

\bibitem{10.1145/3605573.3605616}
G.~Katevenis, M.~Ploumidis, M.~Marazakis,
  \href{https://doi.org/10.1145/3605573.3605616}{Impact of cache coherence on
  the performance of shared-memory based mpi primitives: A case study for
  broadcast on intel xeon scalable processors}, Proceedings of the 52nd
  International Conference on Parallel Processing (2023) 295–305\href
  {http://dx.doi.org/10.1145/3605573.3605616}
  {\path{doi:10.1145/3605573.3605616}}.
\newline\urlprefix\url{https://doi.org/10.1145/3605573.3605616}

\bibitem{SZUSTAK2023623}
L.~Szustak, M.~Lawenda, S.~Arming, G.~Bankhamer, C.~Schweimer, R.~Elsässer,
  \href{https://www.sciencedirect.com/science/article/pii/S0167739X23002571}{Profiling
  and optimization of python-based social sciences applications on hpc systems
  by means of task and data parallelism}, Future Generation Computer Systems
  148 (2023) 623--635.
\newblock \href
  {http://dx.doi.org/https://doi.org/10.1016/j.future.2023.07.005}
  {\path{doi:https://doi.org/10.1016/j.future.2023.07.005}}.
\newline\urlprefix\url{https://www.sciencedirect.com/science/article/pii/S0167739X23002571}

\bibitem{10.1145/3468267.3470615}
A.~Yildirim, C.~A. Mader, J.~R. R.~A. Martins,
  \href{https://doi.org/10.1145/3468267.3470615}{Accelerating parallel cfd
  codes on modern vector processors using blockettes}, Proceedings of the
  Platform for Advanced Scientific Computing Conference\href
  {http://dx.doi.org/10.1145/3468267.3470615}
  {\path{doi:10.1145/3468267.3470615}}.
\newline\urlprefix\url{https://doi.org/10.1145/3468267.3470615}

\bibitem{AMD4thGenSpec}
\href{https://www.amd.com/en/campaigns/epyc-9004-architecture}{{4th gen AMD
  EPYC Processors Architecture (white-paper)}} (2023).
\newline\urlprefix\url{https://www.amd.com/en/campaigns/epyc-9004-architecture}

\bibitem{SZU20_TPDS}
L.~Szustak, R.~Wyrzykowski, O.~T., V.~Mele, {Correlation of Performance
  Optimizations and Energy Consumption for Stencil-Based Application on Intel
  Xeon Scalable Processors}, IEEE Trans. Parallel Distrib. Syst. 31~(11) (2020)
  2582 -- 2593.

\bibitem{SZU21_TPDS}
L.~Szustak, et~al., {Architectural Adaptation and Performance-Energy
  Optimization for CFD Application on AMD EPYC Rome}, IEEE Trans. Parallel
  Distrib. Syst. 32~(12) (2021) 2852 -- 2866.

\bibitem{AMDEPYC2ndGen}
\href{https://www.amd.com/en/processors/epyc-7002-series}{{AMD EPYC™ 7002
  Series Processors}}, {Accessed}: 2024-05-17 (2024).
\newline\urlprefix\url{https://www.amd.com/en/processors/epyc-7002-series}

\bibitem{IOConHub}
\href{https://en.wikipedia.org/wiki/I/O_Controller_Hub}{{I/O Controller Hub}},
  {Accessed}: 2024-05-17 (2024).
\newline\urlprefix\url{https://en.wikipedia.org/wiki/I/O_Controller_Hub}

\bibitem{Scotchlib}
\href{https://www.labri.fr/perso/pelegrin/scotch/}{{SCOTCH library website}},
  {Accessed}: 2024-05-17 (2024).
\newline\urlprefix\url{https://www.labri.fr/perso/pelegrin/scotch/}

\bibitem{HiDALGO2}
\href{https://www.hidalgo2.eu/}{{HiDALGO2 project website}}, {Accessed}:
  2024-05-17 (2024).
\newline\urlprefix\url{https://www.hidalgo2.eu/}

\bibitem{OpenFOAMStanSolv}
\href{https://www.openfoam.com/documentation/user-guide/a-reference/a.1-standard-solvers}{{OpenFOAM
  Standard solvers}}, {Accessed}: 2024-05-17 (2024).
\newline\urlprefix\url{https://www.openfoam.com/documentation/user-guide/a-reference/a.1-standard-solvers}

\bibitem{Polytope}
M.~Leuridan, J.~Hawkes, T.~Quintino,
  \href{https://meetingorganizer.copernicus.org/EGU23/EGU23-8839.html}{Polytope:
  Feature extraction for improved access to petabyte-scale datacubes},
  EGU23-8839\href
  {http://dx.doi.org/https://doi.org/10.5194/egusphere-egu23-8839}
  {\path{doi:https://doi.org/10.5194/egusphere-egu23-8839}}.
\newline\urlprefix\url{https://meetingorganizer.copernicus.org/EGU23/EGU23-8839.html}

\bibitem{Copert}
\href{https://copert.emisia.com/}{{COPERT - EU standard vehicle emissions
  calculator}}, {Accessed}: 2024-05-17 (2024).
\newline\urlprefix\url{https://copert.emisia.com/}

\bibitem{AOCC}
\href{https://www.amd.com/en/developer/aocc.html}{{AMD Optimizing C/C++ and
  Fortran Compilers (AOCC)}} (2023).
\newline\urlprefix\url{https://www.amd.com/en/developer/aocc.html}

\bibitem{Galeazzo2024}
F.~C.~C. Galeazzo, R.~G. Wei{\ss}, S.~Lesnik, H.~Rusche, A.~Ruopp,
  \href{https://doi.org/10.20944/preprints202404.0219.v1}{Understanding
  superlinear speedup in current hpc architectures}, Preprints\href
  {http://dx.doi.org/10.20944/preprints202404.0219.v1}
  {\path{doi:10.20944/preprints202404.0219.v1}}.
\newline\urlprefix\url{https://doi.org/10.20944/preprints202404.0219.v1}

\bibitem{uProf41}
\href{https://www.amd.com/en/developer/uprof.html}{{AMD uProf User Guide 4.1}}.
\newline\urlprefix\url{https://www.amd.com/en/developer/uprof.html}

\bibitem{Pearson}
\href{https://en.wikipedia.org/wiki/Pearson_correlation_coefficient}{{Pearson
  correlation coefficient}}, {Accessed}: 2024-05-17 (2024).
\newline\urlprefix\url{https://en.wikipedia.org/wiki/Pearson_correlation_coefficient}

\end{thebibliography}

\end{document}